\documentclass[twocolumn,showpacs,nofootinbib,prd,preprintnumbers,amssymb]{revtex4}
\usepackage{latexsym}
\usepackage{graphicx,epsf, epsfig, amssymb}% Include figure files
\def\ii{{\rm i}}
\usepackage[usenames,dvipsnames]{xcolor}

\begin{document}

\title{Quasi-normal modes of superfluid neutron stars}

\author{L.~Gualtieri$^1$, E.~M.~Kantor$^2$, M.~E.~Gusakov$^{2,3}$,
  A.~I.~Chugunov$^2$} \address{$^1$Dipartimento di Fisica, ``Sapienza''
  Universit\`a di Roma \& Sezione INFN Roma1, Piazzale Aldo Moro 5, 00185, Roma,
  Italy} \address{$^2$Ioffe Physical-Technical Institute of the Russian Academy
  of Sciences, Polytekhnicheskaya 26, 194021 Saint-Petersburg, Russia}
\address{$^3$Saint-Petersburg State Polytechnical University, Polytekhnicheskaya
  29, 195251 Saint-Petersburg, Russia}

%%%%%%%%%%%%%%%%%%%%%%%%%%%%%%%%%%%%%%%%%%
\begin{abstract} 
  We study non-radial oscillations of neutron stars with superfluid baryons, in
  a general relativistic framework, including finite temperature effects.  Using
  a perturbative approach, we derive the equations describing stellar
  oscillations, which we solve by numerical integration, employing different
  models of nucleon superfluidity, and determining frequencies and gravitational
  damping times of the quasi-normal modes.  As expected by previous results, we
  find two classes of modes, associated to superfluid and non-superfluid degrees
  of freedom, respectively.  We study the temperature dependence of the modes,
  finding that at specific values of the temperature, the frequencies of the two
  classes of quasi-normal modes show avoided crossings, and their damping times
  become comparable.  We also show that, when the temperature is not close to
  the avoided crossings, the frequencies of the modes can be accurately computed
  by neglecting the coupling between normal and superfluid degrees of freedom.
  Our results have potential implications on the gravitational wave emission
  from neutron stars.
\end{abstract}
%%%%%%%%%%%%%%%%%%%%%%%%%%%%%%%%%%%%%%%%%%

%%%%%%%%%%%%%%%%%%%%%%%%%%%%%%%%%%%%%%
% 04.40.Dg: Relativistic stars: structure, stability, and oscillations
% 97.60.Jd: Neutron stars
% 47.75.+f, Relativistic fluid dynamics
% 47.37.+q  Hydrodynamics, superfluidity
% 97.10.Sj  Oscillations, stellar
% 04.30.-w  Gravitational waves, general relativity
% 26.60.-c: Nuclear matter aspects of neutron stars
% 95.30.Sf  Relativity and gravitation

\maketitle

%%%%%%%%%%%%%%%%%%%%%%%%%%%%%%%%%%%%%%%%%%%%%%%%%%%%%%%%%%%%%%%%%%%%%%%%%%%%%%%%%%%%%%
\section{Introduction}
\label{intro}
%%%%%%%%%%%%%%%%%%%%%%%%%%%%%%%%%%%%%%%%%%%%%%%%%%%%%%%%%%%%%%%%%%%%%%%%%%%%%%%%%%%%%%

When a neutron star (NS) is excited by an external or internal event -- such as
a glitch, a close interaction with an orbital companion, or the gravitational
collapse from which it is born -- it can be set into non-radial oscillations,
emitting gravitational waves (GWs) at the frequencies of its quasi-normal modes
(QNMs).  Such oscillations are damped, due to GW emission and to dissipative
processes.  In some cases -- for instance, in presence of rotation -- unstable
modes can also be present in the spectrum; these modes do not require a specific
excitation mechanism, since small fluctuations in this case grow exponentially.

The QNMs of a NS carry invaluable information on the state and composition of
matter at the extreme densities and pressures prevailing in its core, 
which are still poorly understood 
(see, e.g., \cite{Lattimer:2006xb} and references therein). 
Detection of the gravitational emission from a non-radially oscillating NS 
(by second- or, more likely, third-generation gravitational interferometers 
\cite{Andersson:2009yt,anderssonetal13}) would allow us to measure the
frequencies and damping times of the NS QNMs, 
extracting information on the behaviour of matter in the stellar core
\cite{Andersson:1997rn,Kokkotas:1999mn,Benhar:2004xg}. 
In addition, NS oscillations are probably associated 
to a wide variety of interesting astrophysical phenomena, 
such as quasi-periodic oscillations of the electromagnetic radiation
observed in giant flares of soft gamma repeaters
\cite{Israel:2005av,Strohmayer:2005ks,Strohmayer:2006py,Watts:2006mr}.

It is then not surprising that in the last decades, a huge effort has been
done, on the theoretical side, to model -- in a general relativistic framework --
NS oscillations, taking into account all relevant features of the matter composing the NS. 
For many years, however, most studies neglected an important
feature of NS matter: baryon superfluidity.

Theoretical studies (see, e.g., reviews \cite{yls99,ls01}) show that baryon
matter in NS cores becomes superfluid at $T\lesssim 10^8-10^{10}$~K.  This is
also suggested by astrophysical observations.  For instance, it is difficult to
explain the phenomenon of pulsar glitches without invoking baryon superfluidity
\cite{lrr-2008-10}.  Recent observations of the real-time cooling of the NS in
Cassiopeia A supernova remnant \cite{Heinke:2010cr,Elshamouty:2013nfa} can also
be explained by NS models with superfluid baryons in the core
\cite{syhhp11,Page:2010aw}.

Non-radial oscillations of relativistic stars have been studied since the late
sixties (e.g., \cite{tc67,ld83,dl85,cf91}), but the pioneer works neglected
superfluidity.  Oscillations of superfluid stars, in which different components
of the fluid can have different velocities, were first studied in a Newtonian
framework \cite{1988ApJ...333..880E,1994ApJ...421..689L,1995A&A...303..515L,
  lm00,Andersson:2001bz,Prix:2001xc,Prix:2002fk,YL03a,Wong:2008xa,
  Andersson:2008fg,2009MNRAS.397.1464H,2009MNRAS.396..951P,
  Samuelsson:2009xz,pa11,pa12}, and, more recently, in general relativity
\cite{Comer:1999rs,Andersson:2002jd,Yoshida:2003hc,lac08,cg11,gkcg13}. Most of
these papers assumed vanishing temperature.  However, this approximation is not
justified: although after the first minutes of life, the temperature $T$ of a NS
is much lower than the Fermi energy of neutrons [which allows one to use a
zero-temperature equation of state (EoS)], it can be comparable to the critical
temperature $T_{ci}$ at which baryon species $i$ becomes superfluid.  Therefore,
the temperature $T$ determines the fraction of paired baryons as well as the
size of the superfluid region in the core and thus affects the dynamical
properties of a NS.  As discussed in \cite{cg11,kg11}, the assumption $T=0$ can 
lead to qualitatively incorrect results.

Non-radial oscillations of superfluid NSs in general relativity, taking into
account finite temperature effects, have first been studied in
\cite{cg11,gkcg13}, in the so-called ``decoupled limit'', in which the (small)
coupling between non-superfluid and superfluid degrees of freedom is neglected
(see also \cite{gk11}).  In this paper we do not neglect this coupling.  We
derive the fully coupled equations describing non-radial oscillations of
non-rotating, superfluid NSs, generalizing the equations of Lindblom \&
Detweiler \cite{ld83,dl85} to the case of superfluid nuclear matter.  We also
perform numerical integrations of these equations, finding the frequencies and
the gravitational damping times of the QNMs.  We consider two different models
of baryon superfluidity, which -- in our opinion -- capture the features of the
models presently studied in the literature.  Since our stellar models are
non-rotating (and no other source of instability, such as convection, is
present), we only have stable, damped QNMs.

The paper is organized as follows. In Sec.~\ref{hydro} we briefly review
superfluid finite temperature hydrodynamics. In Sec.~\ref{perturb} we derive and
discuss the perturbation equations, and their numerical implementation.  In
Sec.~\ref{models} we describe the microphysics input and the equilibrium stellar
models adopted in this paper. In Sec.~\ref{results} we show and discuss the
results of our numerical integrations, and in Sec.~\ref{concl} we draw our
conclusions.  The explicit expression of our perturbation equations is given in
the Appendix.

%%%%%%%%%%%%%%%%%%%%%%%%%%%
\section{Superfluid finite-temperature hydrodynamics}
\label{hydro}
%%%%%%%%%%%%%%%%%%%%%%%%%%%
The equations of superfluid relativistic finite-temperature hydrodynamics were
reviewed in many papers, see, e.g., Refs.\ \cite{ga06, gusakov07, kg11, gkcg13}.
Here, following Ref.\ \cite{gkcg13}, we assume that the matter of NS cores
consists of neutrons ($n$), protons ($p$) and electrons ($e$), i.e.,
$npe$-matter; and that when the temperature is small enough (see below),
neutrons are superfluid, and protons are superconducting.

Superfluidity affects the dynamical properties of a fluid, leading to a
possibility of co-existence, without dissipation, of several independent motions
with different velocities \cite{khalatnikov89}.  In particular, superfluid
$npe$-matter is described by the three four-velocities: $u^{\mu}$,
$v^{\mu}_{{\rm s}(n)}$, and $v^{\mu}_{{\rm s}(p)}$, where $u^{\mu}$ is the
velocity of the normal (non-superfluid) liquid component (electrons and
Bogoliubov excitations of neutrons and protons) and $v^\mu_{{\rm s}(i)}$ is the
``superfluid'' velocity of particle species $i=n$ or $p$ (the velocity of
superfluid condensate of species $i$).  In what follows, instead of
$v^{\mu}_{{\rm s}(i)}$, we will use the four-vector $w^{\mu}_{(i)} =
\mu_i[v^{\mu}_{s(i)}-u^{\mu}]$, where $\mu_i$ is the relativistic chemical
potential for particle species $i$.

The existence of two additional velocities in superfluid $npe$-matter modifies
the expressions for neutron and proton conserved current densities, which become
\begin{equation}
j^{\mu}_{(i)} = n_i u^{\mu} + Y_{ik} w^{\mu}_{(k)}
\label{jnp}
\end{equation}
(c.f. with the ordinary expression $j^{\mu}_{(i)}=n_i u^{\mu}$).  At the same
time, the electron current density remains unaffected by superfluidity,
\begin{equation}
j^{\mu}_{(e)}=n_{e} u^{\mu}.
\label{je}
\end{equation}
Here and below, the indexes ($i,\,k,\,l$) refer to particle species; in
particular, indexes $i,\,k$ refer to nucleons ($i,\,k = n\,,p$), $l$ refers to
nucleons and electrons ($l=n,\,p,\,e$); $n_l$ is the number density of the
particle specie $l$. Greek letters ($\mu,\nu,\dots$) refer to spacetime indexes,
and the index $j$ refers to purely spatial indexes. Unless otherwise stated, a
summation is assumed over repeated indexes.  We use geometrized units in which
$G=c=1$.

In Eq.\ (\ref{jnp}), $Y_{ik}$ is the symmetric relativistic entrainment matrix,
which is a generalization of the so-called Andreev-Bashkin matrix $\rho_{ik}$
\cite{ab75,bjk96,gh05,gusakov10} to the relativistic case \cite{gkh09a,gkh09b}.
It was first introduced in Ref.~\cite{ga06} and accurately calculated in Refs.\
\cite{gkh09a,gkh09b, ghk14}.  In the non-relativistic limit this matrix is
related to $\rho_{ik}$ by the condition
\begin{equation}
\rho_{ik}=m_i m_k \, Y_{ik},
\label{rho_Y}
\end{equation}
where $m_i$ is the bare nucleon mass, and there is no sum over the indexes $i,k$.
In the case of a one-component superfluid liquid, $\rho_{ik}$ reduces to the
so-called superfluid density $\rho_{\rm s}$ (see, e.g., \cite{khalatnikov89}).
The (symmetric) matrix $Y_{ik}$ generally depends on the Landau parameters
$F_1^{ik}$ of asymmetric nuclear matter and on the temperature $T$
\cite{gkh09b}.  In beta-equilibrium $Y_{ik}$ can be expressed as a function only
depending on the energy density $\rho$ (or the baryon number density $n_b =
n_n+n_p$) and the combinations $T/T_{{\rm c}{n}}$ and $T/T_{{\rm c}{p}}$,
$Y_{ik}=Y_{ik}(\rho, T/T_{{\rm c}{n}}, T/T_{{\rm c}{p}})$, where $T_{{\rm
    c}{n}}(\rho)$ and $T_{{\rm c}{p}}(\rho)$ are the density-dependent neutron
and proton critical temperatures, respectively.  If, for example, $T > T_{{\rm
    c} {n}}$, then all neutrons are normal and the corresponding matrix elements
$Y_{nk}=Y_{kn}$ vanish.

In what follows we will be interested in low-frequency oscillations of a NS
($p$- and $f$-modes) which are well below the electron and proton plasma
frequencies, and therefore preserve quasi-neutrality, $n_e=n_p$.  For a
non-rotating non-magnetized NS this condition, together with continuity
equations for electrons and protons, implies that $j^{\mu}_{({ p})}=j^{\mu}_{({
    e})}$ or, in view of Eqs.\ (\ref{jnp}) and (\ref{je}), that
\begin{equation}
Y_{pk} w^{\mu}_{(k)}=0, 
\label{quasineutrality}
\end{equation}
that is, 
$w^{\mu}_{({ p})}$ and $w^{\mu}_{({ n})}$ are interrelated.

We introduce two four-vectors which will be useful later,
\begin{equation}
X^{\mu} = \frac{Y_{{ n}k} w^{\mu}_{(k)}}{n_{ b}},
\label{X0}
\end{equation}
which depends on the superfluid degrees of freedom, and
\begin{equation}
U^\mu_{({b})} = u^{\mu}+X^{\mu},
\label{ub}
\end{equation}
which we call ``baryon four-velocity'' (strictly speaking, $U^\mu_{(b)}$ is not
a four-velocity, since $U^{\mu}_{({b})} U_{({ b}) \, \mu}=-1$ only in the
linearized theory; see the footnote 4 in Ref.\ \cite{gkcg13} for more details).
As follows from Eqs.\ (\ref{jnp}), (\ref{je}) and (\ref{quasineutrality}), the
baryon current density $j^{\mu}_{({b})}=j^{\mu}_{({ n})}+j^{\mu}_{({p})}$ is
related to $U^\mu_{({b})}$ by the standard equation
\begin{equation}
j^{\mu}_{({ b})}=n_{ b} \,
U^{\mu}_{({ b})},
\label{jb}
\end{equation}
while $j_{ (e)}^{\mu}$ equals
\begin{equation}
j^{\mu}_{ (e)}=n_{e}  \left[U_{({ b})}^{\mu} - X^{\mu} \right].
\label{je2}
\end{equation}
Together with the quasi-neutrality condition ($n_{ e}=n_{p}$) and
Eq. (\ref{quasineutrality}), the equations of superfluid hydrodynamics are
(e.g., Ref.\ \cite{gkcg13}):
\begin{itemize}
\item[($i$)] Continuity equations for baryons ($b$) and electrons ($e$),
\begin{eqnarray}
j^{\mu}_{ ({ b}) ; \, \mu} &=& 0, 
\label{cont_b}\\
j^{\mu}_{({ e}) ; \, \mu} &=& 0.
\label{cont_e}
\end{eqnarray}

\item[($ii$)] Energy-momentum conservation
\begin{equation}
T^{\mu \nu}_{; \,\mu} = 0, 
\label{Tmunu_cons}
\end{equation}
where the stress-energy tensor for the superfluid described above is:
\begin{eqnarray}
T^{\mu \nu} &=& (P+\rho) \, u^{\mu} u^{\nu} + P g^{\mu \nu} 
\nonumber\\
&+& Y_{ik} \left( w^{\mu}_{(i)} w^{\nu}_{(k)} + \mu_i \, w^{\mu}_{(k)} u^{\nu} 
+ \mu_k \, w^{\nu}_{(i)} u^{\mu} \right).
\label{Tmunu}
\end{eqnarray}
\item[($iii$)] Potentiality condition for superfluid motion of neutrons
\begin{eqnarray}
\partial_{\nu} \left[ w_{({n}) \mu} 
+ \mu_{ n} u_{\mu} \right]
&=& \partial_{\mu} \left[ w_{({ n}) \nu} 
+\mu_{ n} u_{\nu} \right]\,.
\label{wmu22}
\end{eqnarray}
\item[($iv$)] The second law of thermodynamics 
\begin{equation}
d \rho =  T \,  d S + \mu_l \, d n_l
+ \frac{Y_{ik}}{2} \,  d \left( w^{\alpha}_{(i)} w_{(k) \alpha} \right).
\label{2ndlaw}
\end{equation}
\end{itemize}
In formulas (\ref{cont_b})--(\ref{2ndlaw}) $g^{\mu \nu}$ is the metric tensor;
$\partial_\mu \equiv \partial/(\partial x^\mu)$; $\rho$, $S$, and $\mu_{e}$ are
the energy density, entropy density, and relativistic electron chemical
potential, respectively.  Finally, $P$ is the pressure given by
\begin{equation}
P=-\rho+ \mu_{ l} n_{l} + TS.
\label{PPP}
\end{equation}
The equations of superfluid hydrodynamics described above should be supplemented
by two additional conditions on the four-vectors $u^{\mu}$ and
$w^{\mu}_{({n})}$,
\begin{eqnarray}
u_{\mu}u^{\mu}&=&-1,
\label{normalization}\\
u_{\mu} w^{\mu}_{({ n})}&=&0\,,
\label{uw}
\end{eqnarray}
i.e., the normalization condition for the four-velocity $u^{\mu}$, and the
requirement that in the comoving frame, the four-vector $w_{(n)}^\mu$ is purely
spatial.  Then we have, using Eqs.\ (\ref{jnp}), (\ref{Tmunu}),
(\ref{normalization}) and (\ref{uw}), $n_{l}=-u_{\mu} j^{\mu}_{(l)}$ ($l=n$,
$p$, $e$) and $\rho = u_{\mu} u_{\nu} T^{\mu \nu}$.

%%%%%%%%%%%%%%%%%%%%%%%%%%%%%%%%%%%%%%%%%%%%%%%%%%%%%%%%%%%%%%%%%%%%%%%%%%%%%%%%%%%%%%
\section{Perturbations of neutron stars with a superfluid phase}\label{perturb}
%%%%%%%%%%%%%%%%%%%%%%%%%%%%%%%%%%%%%%%%%%%%%%%%%%%%%%%%%%%%%%%%%%%%%%%%%%%%%%%%%%%%%%
The theory of relativistic stellar perturbations has been developed, e.g., in
Refs.\ \cite{tc67,ld83,dl85,cf91}.  It allows one to describe the oscillations
of a relativistic star (such as a NS), and in particular to determine the QNMs
of the star.  We have generalized this theory, originally developed to describe
non-superfluid matter, to include a superfluid phase, which is described within
the approach discussed in Sec.~\ref{hydro}.  We here present the derivation of
this generalization.  Our starting point is the Lindblom \& Detweiler (LD)
formulation of the relativistic theory of stellar perturbations for
non-superfluid stars \cite{ld83,dl85}.  We follow the notation and conventions
of Ref.\ \cite{dl85}.

%%%%%%%%%%%%%%%%%%%%%%%%%%%%%%%%%%%%%%%%%%%%%%%%%%%%%%%%%%%%%%%%%%%%%%%%%%%%%%%%%%%%%%
\subsection{The stationary, spherically symmetric background}
%%%%%%%%%%%%%%%%%%%%%%%%%%%%%%%%%%%%%%%%%%%%%%%%%%%%%%%%%%%%%%%%%%%%%%%%%%%%%%%%%%%%%%
We describe stellar oscillations as linear perturbations of a stationary,
spherically symmetric background, i.e., we expand
\begin{eqnarray}
&&g_{\mu\nu}= g^{(0)}_{\mu\nu}+\delta g_{\mu\nu}\,,
\nonumber\\
&&u^\mu=u^{(0)\,\mu}+\delta u^\mu\,,\nonumber\\
&&U^\mu_{(b)} = U^{(0)\,\mu}_{(b)}+\delta U^\mu_{(b)}.\label{gu}
\end{eqnarray}
The Schwarzschild background metric $g^{(0)}_{\mu\nu}$, in the coordinates
$x^\mu=(t,r,\theta,\phi)$, can be written as
\begin{equation}
(ds^{(0)})^2=-e^\nu dt^2+e^\lambda dr^2 
+ r^2(d\theta^2+\sin^2\theta d\phi^2),\label{metric}
\end{equation}
where $\nu(r)$, $\lambda(r)$ are solutions to Eqs.~(\ref{tov}) [see below]. 
The background normal four-velocity is 
$u^{(0)\,\mu}=(e^{-\nu/2},0,0,0)$.

In a non-rotating and non-magnetized star, the unperturbed velocities of normal
and superfluid liquid components coincide, that is (see, e.g.,
\cite{ga06,kg11,gkcg13})
\begin{equation}
w^{(0)\,\mu}_{(n)}=w^{(0)\,\mu}_{(p)}=X^{(0)\,\mu}=0.
\label{sflcondeq}
\end{equation}
Hence $U^{(0)\,\mu}_{(b)}=u^{(0)\,\mu}=(e^{-\nu/2},0,0,0)$ [see Eq.\
(\ref{ub})], while the background stress-energy tensor has exactly the same form
as for a perfect fluid in the absence of superfluidity:
$T^{(0)}_{\mu\nu}=(\rho^{(0)}+P^{(0)})u^{(0)}_\mu u^{(0)}_\nu+P^{(0)}
g^{(0)}_{\mu\nu}$. In other words, the stationary configuration, i.e., the
structure equations and the background spacetime metric, are not affected by
superfluidity (see Ref.\ \cite{ga06} for a detailed discussion of this issue).

Therefore, the hydrostatic structure of a superfluid non-rotating NS is
determined by the solution of standard Tolman-Oppenheimer-Volkov (TOV) equations
[hereafter, a prime denotes differentiation with respect to the radial
coordinate $r$ and the notation $Z^{(0)}$ denotes the value of
some generic quantity $Z$ (e.g., $P$, $\rho$, etc.) 
in the unperturbed star],
\begin{eqnarray}
\nu'&=&\frac{2e^\lambda}{r^2}(m+4\pi P^{(0)}r^3),\nonumber\\
m'&=&4\pi \rho^{(0)} r^2,\nonumber\\
P^{(0)\,\prime}&=&-\frac{1}{2}(\rho^{(0)}+P^{(0)})\nu'\label{tov}\,,
\end{eqnarray}
where $m(r)=r(1-e^{-\lambda})/2$ is the gravitational mass inside the radius
$r$.  If the equation of state (EoS), providing a relation between $P^{(0)}$ and
$\rho^{(0)}$, is known, the TOV equations allow to compute the gravitational
mass $M$ and the circumferential radius $R$ of the NS.

In general, the EoS has the form $P=P(n_b,x_S,x_l)$, where $x_S$ is the entropy
per baryon, and $x_l=n_l/n_b$ are the chemical fractions of different species.
Notice that since the stellar temperature is much smaller than the chemical
potentials, thermal effects do not affect the EoS (but they are relevant for the
transition to the superfluid phase).  Similarly, superfluidity of baryons in NS
interiors does not significantly affect the EoS, because the superfluid
energy gaps are negligible in comparison with the chemical potentials.
Moreover, as discussed in Section \ref{hydro}, we consider $npe$-matter assuming
charge quasi-neutrality ($n_p=n_e$).  Therefore, the fluid can be described by a
two-parameter EoS, say $P=P(n_b,x_e)$.

%%%%%%%%%%%%%%%%%%%%%%%%%%%%%%%%%%%%%%%%%%%%%%%%%%%%%%%%%%%%%%%%%%%%%%%%%%%%%%%%%%%%%%
\subsection{Linear perturbations and harmonic expansion}
%%%%%%%%%%%%%%%%%%%%%%%%%%%%%%%%%%%%%%%%%%%%%%%%%%%%%%%%%%%%%%%%%%%%%%%%%%%%%%%%%%%%%%
We shall only consider perturbations with polar parity, with harmonic index
$l\geq 2$.  The perturbations of the spacetime metric are expanded in tensor
spherical harmonics, and Fourier transformed, as follows:
\begin{eqnarray}
\delta g_{\mu\nu}dx^\mu dx^\nu&=&-\left[e^\nu H^{lm}_0dt^2+2\ii\omega 
rH^{lm}_1dtdr\right.\nonumber\\
&& +e^\lambda H^{lm}_2dr^2
+r^{2}K^{lm}(d\theta^2\nonumber\\
&&\left.+\sin^2\theta d\phi^2)\right]r^lY^{lm}e^{\ii\omega t},\label{expang}
\end{eqnarray} 
where $Y^{lm}(\theta,\phi)$ are the usual spherical harmonics (not to be
confused with the entrainment matrix $Y_{ik}$), while the functions $H^{lm}_0$,
$H^{lm}_1$, $H^{lm}_2$, $K^{lm}$ depend on $r$ only.

%%%%%%%%%%%%%%%%%%%%%%%%%%%%%%%%%%%%%%%%%%%%%%%%%%%%%%%%%%%%%%%%%
\subsubsection{Non-superfluid phase}
%%%%%%%%%%%%%%%%%%%%%%%%%%%%%%%%%%%%%%%%%%%%%%%%%%%%%%%%%%%%%%%%%
The perturbations of the fluid four-velocity are expanded in tensor spherical
harmonics, and Fourier transformed, as:
\begin{eqnarray}
\delta u^\mu&=&\left(\delta u^0,
\ii\omega e^{-(\nu+\lambda)/2}r^{l-1}W^{lm}Y^{lm},\right.\nonumber\\
&&-\ii\omega e^{-\nu/2}r^{l-2}V^{lm}Y^{lm}_{,\theta},
\nonumber\\
&&\left.-\ii\omega e^{-\nu/2}r^{l-2}\sin^{-2}\theta V^{lm} 
Y^{lm}_{,\phi}\right)\label{expanu}{\rm e}^{\ii\omega t}\,,
\end{eqnarray}
where the functions $W^{lm}$ and $V^{lm}$ depend on $r$ only. In addition, if we
denote a generic (scalar) fluid quantity (such as $P$, $\rho$, $n_b$, etc.) as
$Z$, its Eulerian perturbation $\delta Z$ is decomposed as
\begin{equation}
\delta Z=\delta Z^{lm}r^lY^{lm}e^{\ii\omega t}\,.\label{expanz}
\end{equation}
The Lagrangian perturbation $\Delta Z$ (i.e., the perturbation of a given fluid
element) is related to $\delta Z$ by
\begin{equation}
\Delta Z=\delta Z+\xi^j Z_{,j},
\label{lageul}
\end{equation}
where $\xi^j$ is the Lagrangian displacement of the fluid element, related to
the four-velocity perturbation by $\delta u^j=u^{\mu}\xi^j_{,\mu}=\ii\omega
e^{-\nu/2}\xi^j$ (recall that the index $j$ denotes space components,
i.e. $j=1,2,3$).  We note that, in the linear approximation, $u^\mu
Z_{,\mu}=i\omega e^{-\nu/2}(\delta Z+\xi^j Z^{(0)}_{,j}) =i\omega
e^{-\nu/2}\Delta Z$.  The Lagrangian perturbation of the quantity $Z$ is
expanded as $\Delta Z=\Delta Z^{lm}r^lY^{lm}e^{\ii\omega t}$.  From
Eq.~(\ref{expanu}), we can see that
$\xi^r=e^{-\lambda/2}W^{lm}r^{l-1}Y^{lm}e^{\ii\omega t}$, therefore
\begin{equation}
\Delta Z^{lm}=\delta Z^{lm}+Z^{(0)\,\prime}
\frac{e^{-\lambda/2}}{r}W^{lm}\,.\label{lageulharm}
\end{equation}
%

%%%%%%%%%%%%%%%%%%%%%%%%%%%%%%%%%%%%%%%%%%%%%%%%%%%
\subsubsection{Superfluid phase}
%%%%%%%%%%%%%%%%%%%%%%%%%%%%%%%%%%%%%%%%%%%%%%%%%%%
As discussed in Sec.~\ref{hydro}, in a superfluid phase we introduce the
four-vector $U^\mu_{(b)}=u^\mu+X^\mu$, which we call ``baryon four-velocity'',
satisfying (at first order in the perturbations) $U^\mu_{(b)}U_{(b)\,\mu}=-1$.
In equilibrium $U^{(0)\,\mu}_{({b})}=u^{(0)\,\mu}=(e^{-\nu/2},0,0,0)$ and its
perturbation $\delta U^\mu_{(b)}$ has the same expansion as $\delta u^\mu$,
\begin{eqnarray}
\delta U_{(b)}^\mu&=&\left(\delta U_{(b)}^0,
\ii\omega e^{-(\nu+\lambda)/2}r^{l-1}W_{(b)}^{lm}Y^{lm},\right.\nonumber\\
&&-\ii\omega e^{-\nu/2}r^{l-2}V_{(b)}^{lm}Y^{lm}_{,\theta},
\nonumber\\
&&\left.-\ii\omega e^{-\nu/2}r^{l-2}\sin^{-2}\theta V_{(b)}^{lm}
Y^{lm}_{,\phi}\right){\rm e}^{\ii\omega t},
\label{expanUb}
\end{eqnarray}
in terms of the perturbation functions $W_{(b)}^{lm}(r)$ and $V_{(b)}^{lm}(r)$.

In a superfluid phase, we define Lagrangian perturbations in terms of the baryon
four-velocity; therefore, if $Z$ is a generic fluid quantity, $U_{(b)}^\mu
Z_{,\mu}=i\omega e^{-\nu/2}\Delta Z$.
%%%%%%%%%%%%%%%%%%%%%%%%%%%%%%%%%%%%%%%%%%%%%%%%%%%%%%%%%%%%%%%%%%%%%%%%%%%%%%%%%%%%%%
\subsection{Pulsation energy}\label{Emech}
%%%%%%%%%%%%%%%%%%%%%%%%%%%%%%%%%%%%%%%%%%%%%%%%%%%%%%%%%%%%%%%%%%%%%%%%%%%%%%%%%%%%%%
The mechanical energy stored in a QNM with frequency $\omega=\sigma+{\rm
  i}/\tau_{\rm GW}$ can be evaluated, as discussed in \cite{gkcg13}, in terms of
the eigenfunctions of the mode and can be split into two terms:
\begin{equation}
E_{\rm mech}=E_{{\rm mech}\,(b)}+E_{\rm mech\,(sfl)},
\label{Emech1}
\end{equation}
where (see Eqs.\ (72) and (73) of Ref.\ \cite{gkcg13})
\begin{eqnarray}
E_{{\rm mech}\,(b)}(t)&=&e^{-2t/\tau_{\rm GW}}\frac{\sigma^2}{2}\int_0^R(P^{(0)}+\rho^{(0)})
e^{(\lambda-\nu)/2}r^{2l}\nonumber\\
&&\left[|W^{lm}_{(b)}|^2+l(l+1)|V^{lm}_{(b)}|^2\right]dr, 
\label{Emechb}\\
E_{\rm mech\,(sfl)}(t)&=&e^{-2t/\tau_{\rm GW}}\frac{\sigma^2}{2}\int_0^R(P^{(0)}+\rho^{(0)})
e^{(\lambda-\nu)/2}r^{2l}\nonumber\\
&& y \, \left[|W^{lm}_{\rm(sfl)}|^2+l(l+1)|V^{lm}_{\rm(sfl)}|^2\right] dr, 
\label{Emechsfl}
\end{eqnarray}
and
\begin{equation}
y \equiv \frac{n_b Y_{pp}}{\mu_{n}(Y_{nn}Y_{pp}-Y_{np}^2)}-1.
\label{y}
\end{equation}
In Eq.\ (\ref{Emechsfl}) the perturbation functions $W^{lm}_{\rm(sfl)}$ and
$V^{lm}_{\rm(sfl)}$ represent the harmonic components of the four-vector $X^\mu$
defined in Eq.~(\ref{X0}):
\begin{eqnarray}
X^\mu&=&\left(0,
\ii\omega e^{-(\nu+\lambda)/2}r^{l-1}W_{\rm(sfl)}^{lm}Y^{lm},\right.\nonumber\\
&&-\ii\omega e^{-\nu/2}r^{l-2}V_{\rm(sfl)}^{lm}Y^{lm}_{,\theta},
\nonumber\\
&&\left.-\ii\omega e^{-\nu/2}r^{l-2}\sin^{-2}\!\theta V_{\rm(sfl)}^{lm}
Y^{lm}_{,\phi}\right){\rm e}^{\ii\omega t}\label{expanX}.
\end{eqnarray}
Vanishing of $X^0$ follows from Eqs.\ (\ref{quasineutrality}), (\ref{X0}),
(\ref{uw}), and (\ref{sflcondeq}) (see also Eq.~(42) of \cite{gkcg13}). We note
that Eq.~(\ref{ub}) implies
\begin{eqnarray} 
W^{lm}_{\rm(sfl)}&=&W^{lm}_{(b)}-W^{lm},\nonumber\\ 
V^{lm}_{\rm(sfl)}&=&V^{lm}_{(b)}-V^{lm}\,.\label{decdegr}
\end{eqnarray} 
The four-vector $X^\mu$ -- and thus its harmonic components $W^{lm}_{\rm(sfl)}$,
$V^{lm}_{\rm(sfl)}$ -- is associated to superfluid degrees of freedom.  If
$X^\mu=0$ then superfluid degrees of freedom are not excited and superfluid and
normal liquid components move with the same velocity (the so-called
``co-moving'' oscillations, similar to those of a non-superfluid matter)
\footnote{Here and in what follows by ``superfluid'' we, by definition,
  understand degrees of freedom associated with the vector $X^{\mu}$, whose
  spatial components depend on the difference between the normal and superfluid
  velocities (see Sec.\ \ref{hydro}).  Correspondingly, by ``normal'' we imply
  degrees of freedom associated with the baryon four-velocity $U^{\mu}_{(b)}$.
  We remark that this is only a convention, even though in Sec. \ref{spect} we
  will justify this choice {\it a posteriori}.}.
In this case only the first term, $E_{{\rm mech}\,(b)}$,  survives in Eq. (\ref{Emech1}).
%%%%%%%%%%%%%%%%%%%%%%%%%%%%%%%%%%%%%%%%%%%%%%%%%%%%%%%%%%%%%%%%%%%%%%%%%%%%%%%%%%%%%%
\subsection{The Lindblom \& Detweiler equations (non-superfluid matter)}\label{LD}
%%%%%%%%%%%%%%%%%%%%%%%%%%%%%%%%%%%%%%%%%%%%%%%%%%%%%%%%%%%%%%%%%%%%%%%%%%%%%%%%%%%%%%
We here briefly discuss the derivation of the LD equations in the case of a NS
composed of non-superfluid matter.

If we substitute the expansions (\ref{expang}), (\ref{expanu}), and
(\ref{expanz}) into the linearized Einstein's equations
\begin{equation}
\delta G_{\mu\nu}=8\pi \delta T_{\mu\nu}\label{leq}
\end{equation} 
where
\begin{eqnarray}
\delta T_{\mu\nu}&=&(\delta\rho+\delta P) u^{(0)}_\mu 
u^{(0)}_\nu+(\rho^{(0)}+P^{(0)})(u^{(0)}_\mu\delta u_\nu\nonumber\\
&&+u^{(0)}_\nu\delta u_\mu)+P^{(0)}\delta g_{\mu\nu}+\delta P g^{(0)}_{\mu\nu}\,,\label{dtmn}
\end{eqnarray} 
we get a set of equations for the seven perturbation functions
$H^{lm}_0,H^{lm}_1,K^{lm},W^{lm},V^{lm}$, $\delta\rho^{lm},\delta P^{lm}$
($H_2^{lm}$ does not appear explicitly in the equations because Einstein's
equations imply $H_2^{lm}=H_0^{lm}$ for $l\geq 2$).  In the LD formulation
\cite{dl85}, there are four first-order differential equations [equations
(8)--(11) of Ref.\ \cite{dl85}] and two algebraic relations [equations (5) and
(6) of Ref.\ \cite{dl85}].  In addition, $\delta\rho^{lm}$ and $\delta P^{lm}$
are related by the EoS, since
\begin{equation}
\Delta P^{lm} =c_s^2\Delta\rho^{lm}
\label{dpde}
\end{equation} 
where 
\begin{equation}
c_s^2\equiv\left(\frac{\partial P}{\partial\rho}\right)^{(0)}_{x_e}\,.
\end{equation}
This is due to the fact that we consider non-dissipative hydrodynamics,
therefore in the oscillation timescale $\Delta x_e=0$, and the term $(\partial
P/\partial x_e)_{\rho}\Delta x_e$ in Eq.~(\ref{dpde}) vanishes.  The property
$\Delta x_e=0$ can be shown, for instance, as follows.  Continuity equation for
baryons [$(n_bu^\mu)_{;\mu}=0$] can be written as
\begin{equation}
u^\mu n_{b,\mu}+n_bu^\mu_{~;\mu}=\ii\omega e^{-\nu/2}\Delta n_b+n_bu^\mu_{~;\mu}=0\,.
\label{cont2}
\end{equation}
Since continuity equation for electrons [$(n_e u^\mu)_{;\mu}=0$] also holds, we
have that $\ii\omega e^{-\nu/2}\Delta n_e+n_eu^\mu_{~;\mu}=0$, which, compared
with (\ref{cont2}), yields $\Delta n_b/n_b=\Delta n_e/n_e$, and then $\Delta
x_e=0$.  We remark that Eq.\ (\ref{dpde}) is equivalent to
\begin{equation}
\Delta P^{lm}=\frac{\gamma P^{(0)}}{n_b^{(0)}}\Delta n_b^{lm}\,,\label{dpdn}
\end{equation}
where
\begin{equation}
\gamma\equiv\frac{n_b^{(0)}}{P^{(0)}}\left(\frac{\partial P}{\partial n_b}\right)^{(0)}_{x_e}\,.
\end{equation} 
Eq.~(\ref{dpde}) [or, equivalently, (\ref{dpdn})] allows one to reduce the
number of perturbation functions to six:
\begin{equation}
H_0^{lm},H_1^{lm},K^{lm},W^{lm},V^{lm},X^{lm},
\end{equation}
where we have defined 
\begin{equation}
X^{lm}\equiv-e^{\nu/2}\Delta P^{lm}=-e^{\nu/2}\delta P^{lm}-P^{(0)\,\prime}
\frac{e^{(\nu-\lambda)/2}}{r}W^{lm}\label{defX}
\end{equation}
(not to be confused with the four-vector $X^{\mu}$ introduced in the previous
Section).  Therefore, in the non-superfluid case the LD equations are fully
determined, because they are six (differential or algebraic) equations for six
perturbation functions.

The relation between $\delta P^{lm}$ and $\delta\rho^{lm}$ (or
between $\delta P^{lm}$ and $\delta n_b^{lm}$) only affects one of the LD
equations, namely
\begin{eqnarray}
W^{lm\,\prime}&=&-\frac{l+1}{r}W^{lm}+re^{\lambda/2}
\left[\frac{e^{-\nu/2}}{\gamma P^{(0)}}X^{lm}\right.\nonumber\\
&&\left.-\frac{l(l+1)}{r^2}V^{lm}+\frac{1}{2}H_0^{lm}+K^{lm}\right].\label{eq10b}
\end{eqnarray}
To show how this occurs let us note that Eq.~(\ref{eq10b}) can be derived from
the continuity equation for baryons, which in our perturbative scheme can be
written in the form (\ref{cont2}).  Substitution of the perturbative expansion
for the four-velocity (\ref{expanu}) into Eq.~(\ref{cont2}) yields
\begin{eqnarray} 
W^{lm\,\prime}&=&-\frac{l+1}{r}W^{lm}+re^{\lambda/2}\left[-\frac{\Delta
n_b^{lm}}{n_b^{(0)}}\right.\nonumber\\
&&\left.-\frac{l(l+1)}{r^2}V^{lm}+\frac{1}{2}H_0^{lm}+K^{lm}\right]\,.\label{eq10a}
\end{eqnarray}
Then, using Eq.~(\ref{dpdn}) together with the definition (\ref{defX}), we
obtain Eq.~(\ref{eq10b}), that is, Eq.~(10) in the article by Detweiler \&
Lindblom \cite{dl85}.
%%%%%%%%%%%%%%%%%%%%%%%%%%%%%%%%%%%%%%%%%%%%%%%%%%%%%%%%%%%%%%%%%%%%%%%%%%%%%%%%%%%%%%
\subsection{Perturbation equations for a superfluid star}
\label{gLD}
%%%%%%%%%%%%%%%%%%%%%%%%%%%%%%%%%%%%%%%%%%%%%%%%%%%%%%%%%%%%%%%%%%%%%%%%%%%%%%%%%%%%%%

A remarkable property of the formulation of Refs.~\cite{gk11,gkcg13} is that in
a superfluid phase the stress-energy tensor perturbation has formally the same
form as in a non-superfluid phase [see Eq.~(\ref{dtmn})], with $\delta u^\mu$
replaced by $\delta U^\mu_{(b)}$:
\begin{eqnarray}
\delta T_{\mu\nu}&=&\left(\delta\rho+\delta P\right) U^{(0)}_{(b)\,\mu}
U^{(0)}_{(b)\,\nu}+\left(\rho^{(0)}+P^{(0)}\right) 
\left[U^{(0)}_{(b)\,\mu}\delta U_{(b)\,\nu} \right.\nonumber\\
&&\left.+U^{(0)}_{(b)\,\nu}\delta U_{(b)\mu}\right]
+P^{(0)}\delta g_{\mu\nu}+\delta P g^{(0)}_{\mu\nu}\,.\label{dtmn2}
\end{eqnarray}
Therefore, the perturbation equations have formally the same expressions as the
LD equations, with $W^{lm}$ and $V^{lm}$ [the variables of the expansion
(\ref{expanu})] replaced, respectively, by $W_{(b)}^{lm}$ and $V_{(b)}^{lm}$
[the variables of the expansion (\ref{expanUb})].  The only exception is
equation (10) of Ref.\ \cite{dl85} [i.e., Eq.~(\ref{eq10b}) of this paper],
which was derived by making use of the relation (\ref{dpdn}), not valid in
superfluid matter (see below).

As we have noted in Sec.~\ref{LD}, Eq.~(\ref{eq10b}) follows from the
perturbative expansion of the baryon continuity equation and from
Eq.~(\ref{dpdn}), which relates $\Delta P^{lm}$ and $\Delta n_b^{lm}$.  We shall
now determine how these equations are modified in the superfluid case.

As discussed in Sec.\ \ref{hydro}, the baryon current density $j^\mu_{(b)}=n_b
U_{(b)}^\mu$ and the electron current density
$j^\mu_{(e)}=n_e[U^\mu_{(b)}-X^\mu]$ satisfy the continuity equations
$j^\mu_{(b)~;\mu}=j^\mu_{(e)~;\mu}=0$.  This implies
\begin{eqnarray}
\left[n_b U_{(b)}^\mu\right]_{;\mu}&=&0,   
\label{contn}\\
\left[n_e U_{(b)}^\mu\right]_{;\mu}&=&(n_e X^\mu)_{;\mu}\,.\label{conte}
\end{eqnarray}
Following \cite{gkcg13}, we define
\begin{equation}
\delta n_{e\,({\rm sfl})}\equiv \frac{e^{\nu/2}}{\ii\omega}(n_e X^\mu)_{;\mu}\,.
\end{equation}
The continuity equation for baryons (\ref{contn}) has the same form as in the
non-superfluid case [with $u^\mu$ replaced by $U^\mu_{(b)}$], and its
perturbative expansion yields [see Eq.~(\ref{eq10a})]
\begin{eqnarray}
W_{(b)}^{lm\,\prime}&=&-\frac{l+1}{r}W^{lm}_{(b)}+re^{\lambda/2}\left[-\frac{\Delta
n_b^{lm}}{n_b^{(0)}}\right.\nonumber\\
&&\left.-\frac{l(l+1)}{r^2}V^{lm}_{(b)}+\frac{1}{2}H^{lm}_0+K^{lm}\right]\,.
\end{eqnarray}  
The continuity equation for electrons (\ref{conte}), instead, is different from
that in the non-superfluid case, and gives
\begin{equation} 
\frac{\Delta n_e}{n^{(0)}_e}-\frac{\delta n_{e({\rm sfl})}}{n^{(0)}_e} 
=\frac{ie^{\nu/2}}{\omega}U_{(b)~;\mu}^\mu=
\frac{\Delta n_b}{n_b^{(0)}}\,.
\end{equation}
This implies that (expanding $\delta n_{e({\rm sfl})}=\delta n_{e({\rm
    sfl})}^{lm}r^lY_{lm}e^{\ii\omega t}$)\,,
\begin{equation}
n_b^{(0)}\Delta x^{lm}_e=\Delta n^{lm}_e-x_e^{(0)}
\Delta n_b^{lm}=\delta n^{lm}_{e({\rm sfl})}\,.
\end{equation}
In the superfluid case, then, $\Delta x_e^{lm}$ does not vanish, 
and Eq.~(\ref{dpdn}) is replaced by
\begin{equation}
\Delta P^{lm}=\frac{\gamma P^{(0)}}{n_b^{(0)}}\Delta n_b^{lm}+
\left(\frac{\partial P}{\partial x_e}\right)^{(0)}_{n_b}
\frac{\delta n^{lm}_{e({\rm sfl})}}{n_b^{(0)}}\label{dpdn2}\,.
\end{equation}
Therefore
\begin{eqnarray}
 -\frac{\Delta n_b^{lm}}{n_b^{(0)}}&=&
-\frac{\Delta P^{lm}}{\gamma P^{(0)}}+\frac{1}{\gamma
 P^{(0)}n_b^{(0)}}\left(\frac{\partial P}{\partial x_e}
\right)^{(0)}_{n_b}\delta n_{e({\rm sfl})}^{lm}\nonumber\\
&=&\frac{e^{-\nu/2}}{\gamma P^{(0)}}{\cal X}^{lm},
\end{eqnarray}
where we have defined
\begin{equation}
{\cal X}^{lm}\equiv X^{lm}+e^{\nu/2}\left(\frac{\partial P}{\partial n_e}
\right)^{(0)}_{n_b}\delta n^{lm}_{e({\rm sfl})}\,.
\label{defcalX}
\end{equation}
Eq.~(\ref{eq10a}) is then replaced, in the superfluid case, by
\begin{eqnarray} 
W^{lm\,\prime}_{(b)}&=&-\frac{l+1}{r}W^{lm}_{(b)}+
re^{\lambda/2}\left[\frac{e^{-\nu/2}}{\gamma P^{(0)}}{\cal X}^{lm}
\right.\nonumber\\
&&\left.-\frac{l(l+1)}{r^2}V^{lm}_{(b)}+\frac{1}{2}H^{lm}_0
+K^{lm}\right]\,.\label{eq10d}
\end{eqnarray}
The new set of equations depends on a new perturbation quantity, $\delta
n_{e({\rm sfl})}^{lm}$, which, as shown in Ref.\ \cite{gkcg13}, can be expressed
in terms of the redshifted chemical potential imbalance $\delta\mu^\infty$,
defined as:
\begin{equation}
\delta\mu^\infty = e^{\nu/2} \delta\mu \equiv e^{\nu/2}(\mu_n-\mu_p-\mu_e)\,.
\label{defdmuin}
\end{equation}
Note that $\delta\mu^\infty$ is a first order quantity, since it vanishes on the
background.  As discussed in Refs.\ \cite{gk11,cg11,gkcg13}, the chemical
potential imbalance is related to the space components of the four-vector
$X^\mu=U^\mu_{(b)}-u^\mu$ by
\begin{equation}
X_j=\frac{\ii n_e}{\mu_n n_b\omega y}\partial_j(\delta\mu^\infty),
\label{eq90}
\end{equation}
where $y$ is given by Eq.\ (\ref{y}).  Expanding\,\footnote{
  Note that this expansion is slightly different from that in Ref.\
  \cite{gkcg13}; the quantity $\delta\mu^{lm}$ appearing in Ref.\ \cite{gkcg13}
  reads, with our conventions, $r^l\delta\mu^{lm}$.}
$\delta\mu^\infty=\delta\mu^{lm}r^lY_{lm}e^{\ii\omega t}$, we have [cf.\ Eqs.\
(92) and (100) of Ref.\ \cite{gkcg13}]:
\begin{equation}
\delta n^{lm}_{e({\rm sfl})}=\frac{e^{-\nu/2}}{\left(\frac{\partial
\delta\mu}{\partial n_e}\right)^{(0)}_{n_b}}
(\delta\mu^{lm}-\delta\mu_{\rm norm}^{lm}),
\label{exprnsfl}
\end{equation}
where
\begin{equation}
\delta\mu_{\rm norm}^{lm}\equiv e^{\nu/2}n_b^{(0)}\left(\frac{\partial\delta
\mu}{\partial n_b}\right)^{(0)}_{x_e}\beta_1^{lm}
\label{exprmunorm}
\end{equation} 
with
\begin{eqnarray}
\beta_1^{lm}&=&K^{lm}+\frac{1}{2}H_0^{lm}-
\frac{e^{-\lambda/2}}{r}\left(W^{lm\,\prime}_{(b)}
+\frac{l+1}{r}W^{lm}_{(b)}\right)\nonumber\\
&&-\frac{l(l+1)}{r^2}V^{lm}_{(b)}\,.
\end{eqnarray}
Substituting this expression into Eq.~(\ref{eq10d}), we
find
\begin{equation}
\beta_1^{lm}=-\frac{e^{-\nu/2}}{\gamma P^{(0)}}{\cal X}^{lm}
\end{equation}
and from Eqs.~(\ref{defcalX}), (\ref{exprnsfl}), (\ref{exprmunorm}) we finally obtain
\begin{eqnarray}
{\cal X}^{lm}&=&\frac{1}{1-\gamma_2}\left(X^{lm}+\gamma_3n_b^{(0)}
\delta\mu^{lm}\right),\label{deftcx}
\end{eqnarray} 
where
\begin{equation}
\gamma_2\equiv\frac{\left(\frac{\partial P}{\partial n_e}\right)^{(0)}_{n_b}}{\left(
\frac{\partial P}{\partial n_b}\right)^{(0)}_{x_e}}
\frac{\left(\frac{\partial \delta\mu}{\partial n_b}\right)^{(0)}_{x_e}}{\left(\frac{\partial 
\delta\mu}{\partial n_e}\right)^{(0)}_{n_b}}\,,~~~~
\gamma_3\equiv\frac{\left(\frac{\partial P}{\partial n_e}\right)^{(0)}_{n_b}}{n_b\left(
\frac{\partial\delta\mu}{\partial n_e}\right)^{(0)}_{n_b}}\,.
\end{equation}
The perturbation functions ($H_0^{lm}$, $H_1^{lm}$, $K^{lm}$, $W_{(b)}^{lm}$,
$V_{(b)}^{lm}$, $X^{lm}$) are then coupled by Eqs.\ (\ref{eq10d}) to the new
quantity $\delta\mu^{lm}$, describing the superfluid degrees of freedom.

The equation for the perturbation function $\delta\mu^{lm}$ follows from the
energy-momentum conservation (\ref{Tmunu_cons}) and the potentiality condition
(\ref{wmu22}), and was obtained in Ref.\ \cite{gkcg13}.  In the notations of
this article, it can be written as
\begin{eqnarray}
\delta\mu^{lm\,\prime\prime}=&-&\left[\frac{h'}{h}-\frac{\lambda'}{2}
+\frac{2(l+1)}{r}\right]\delta\mu^{lm\,\prime}\nonumber\\
&&-\left[(1-e^\lambda)\frac{l(l+1)}{r^2}+\frac{l}{r}\left(\frac{h'}{h}-
\frac{\lambda'}{2}\right)\right]\delta\mu^{lm}\nonumber\\
&&+e^{\lambda-\nu/2}\frac{\omega^2}{h{\cal B}}\left[\delta\mu^{lm}+
\frac{\gamma_2}{n_b^{(0)}\gamma_3}{{\cal X}}^{lm}\right],\label{ldmu}
%\frac{\gamma_2}{n_b^{(0)}\gamma_3}{\tilde{\cal X}}^{lm}\right],\label{ldmu}
\end{eqnarray} 
where 
\begin{equation}
{\cal B}\equiv\left(\frac{\partial\delta\mu}{\partial n_e}\right)^{(0)}_{n_b}\label{defb};\,\,
h\equiv e^{\nu/2}\frac{n_e^{(0)\,2}}{\mu_n^{(0)}n_b^{(0)}y}.\label{defh}
\end{equation}
We can conclude that, in the case of superfluid matter, the LD equations are
modified as follows:
\begin{itemize}
\item the functions $W^{lm}(r)$ and $V^{lm}(r)$ are replaced with
  $W^{lm}_{(b)}(r)$ and $V^{lm}_{(b)}(r)$, respectively;
\item a new perturbation function $\delta\mu^{lm}$, satisfying Eq.~(\ref{ldmu}),
  is introduced; 
\item Eq.~(10) of Ref.\ \cite{dl85} is replaced by our Eq.~(\ref{eq10d}). Note
%  that the quantity ${\tilde{\cal X}}^{lm}$ only enters in this equation; the
  that the quantity ${{\cal X}}^{lm}$ 
(which is related to $\delta\mu^{lm}$ by Eq.\ (\ref{deftcx}))
only enters in this equation; the 
 other LD equations [Eqs.~(5), (6), (8), (9), (11) of \cite{dl85}] depend on
  $X^{lm}\equiv -e^{\nu/2}\Delta P^{lm}$, and have the same form (with the
  replacement $W^{lm} \rightarrow W^{lm}_{(b)}$ and $V^{lm} \rightarrow
  V^{lm}_{(b)}$) as in the non-superfluid case.
\end{itemize}
The full set of the perturbation equations is summarized in the Appendix.

%%%%%%%%%%%%%%%%%%%%%%%%%%%%%%%%%%%%%%%%%%%%%%%%%%%%%%%%%%%%%%%%%%%%%%%%%%%%%%%%%%%%%%
\subsection{Boundary conditions}\label{bc}
%%%%%%%%%%%%%%%%%%%%%%%%%%%%%%%%%%%%%%%%%%%%%%%%%%%%%%%%%%%%%%%%%%%%%%%%%%%%%%%%%%%%%%
We look for solutions of the perturbation equations describing a star
oscillating in its QNMs.  We here discuss the boundary conditions corresponding
to such solutions.

The boundary conditions depend on the structure of the superfluid phase of
neutrons.  Microscopic calculations predict the so-called bell-shaped profile of
critical temperature (see Sec.~\ref{models}), which has a maximum at a certain
value of the density, and decreases at larger and lower densities (see, e.g.,
Figs.\ \ref{Fig:Tc1} and \ref{Fig:Tc2}).  As a result, depending on the
parameters of the neutron critical temperature profile, stellar model, and
stellar temperature we have two possibilities: two-layer stars or three-layer
stars.\footnote{For simplicity we do not account for superfluidity of neutrons
  in the crust, which could lead to additional layers; we also assume constant
  redshifted stellar temperature.}

In the case of a two-layer star we have a superfluid internal layer
[where $T(r)<T_{{\rm c}n}(r)$] and a non-superfluid external layer (see Fig.\
\ref{Fig:Tc1}, where the dashed region is superfluid at a redshifted temperature
$T^\infty=4\times 10^8$ K).  When the maximum of $T_{{\rm c}n}$ corresponds to a
density which is lower than the central density of the star, configurations with
three layers are possible.  In this case we have non-superfluid internal and
external layers and neutron superfluidity in between (see the dashed region in
Fig.\ \ref{Fig:Tc2}).  We denote the inner and outer radii of the neutron
superfluid phase by $r_i$ and $r_f$, respectively.

We note that when $T_{{\rm c}n}(r)=T(r)$ then $h=0$, as expected from
Eqs.~(\ref{y}) and (\ref{defh}), since $Y_{ni}\rightarrow0$ in this
limit. Therefore, Eq.~(\ref{ldmu}) implies
\begin{eqnarray}
&&\left(\delta\mu^{lm\,\prime}+\frac{l}{r}\delta\mu^{lm}\right)_{T_{{\rm c}n}(r)=T(r)}\nonumber\\
&&=\left[e^{\lambda-\nu/2}\frac{\omega^2}{h'{\cal B}}\left(\delta\mu^{lm}
+\frac{\gamma_2}{n^{(0)}_b\gamma_3}
{{\cal X}^{lm}}\right)\right]_{T_{{\rm c}n}(r)=T(r)}\,.\label{bcrif}
%{\tilde{\cal X}^{lm}}\right)\right]_{T_{{\rm c}n}(r)=T(r)}\,.\label{bcrif}
\end{eqnarray} 
%

%%%%%%%%%%%%%%%%%%%%%%%%%%%%%%%%%%%%%%%%%%%%%%%%%%%%%%%%%%%%%%%%%%%%%%%%%%%%%%%%%%%%%%
\subsubsection{Inner boundary conditions}
%%%%%%%%%%%%%%%%%%%%%%%%%%%%%%%%%%%%%%%%%%%%%%%%%%%%%%%%%%%%%%%%%%%%%%%%%%%%%%%%%%%%%%
To impose boundary conditions at the stellar center we have to consider an
asymptotic expansion of the perturbation equations at $r \rightarrow 0$.  We
expand the perturbation functions as
$X^{lm}(r)=X^{lm}(0)+\frac{1}{2}r^2X^{lm\,\prime\prime}(0)+\dots$,
$H_1^{lm}(r)=H_1^{lm}(0)+\frac{1}{2}r^2H_1^{lm\,\prime\prime}(0)+\dots$, etc.\
and the background quantities as $\rho^{(0)}=\rho_0+\frac{1}{2}r^2\rho_2+\dots$,
$P^{(0)}=P_0+\frac{1}{2}r^2P_2+\dots$, etc., and replace these expressions in
the perturbation equations.
\begin{itemize}
\item If the star has a three-layer structure, neutrons at its center are
  non-superfluid, and the perturbations at the center are described by the LD
  equations \cite{dl85}.  We find at the lowest order \cite{dl85}:
\begin{eqnarray}
 X^{lm}(0)&=&(\rho_0+P_0)e^{\nu_0/2}\left\{\left[\frac{4\pi}{3}
(\rho_0+3P_0)\right.\right.\nonumber\\
&&\left.\left.-\omega^2\frac{e^{-\nu_0}}{l}
   \right]W^{lm}(0)+\frac{1}{2}K^{lm}(0)\right\},\nonumber\\
 H_1^{lm}(0)&=&\frac{2lK^{lm}(0)+16\pi(\rho_0+P_0)
W^{lm}(0)}{l(l+1)}\label{ld1617}\,.
\end{eqnarray}
This lowest order is sufficient to solve the LD equations and find the QNMs with
good accuracy: we do not need to include second order terms in the expansion.

Imposing the boundary conditions (\ref{ld1617}) we have, for each value of
$\omega$, two independent solutions of the LD equations.

We integrate the LD equations up to $r=r_i$, where we require
$W_{(b)}^{lm}(r_{i\,+})=W^{lm}(r_{i\,-})$, continuity of $H_1^{lm}$, $K^{lm}$,
$X^{lm}$, and impose Eq.~(\ref{bcrif}) which allows us to determine
$\delta\mu^{lm}$ up to an arbitrary constant. Therefore, we have three
independent solutions of Eqs.~(\ref{eqH1})--(\ref{eqcX}) satisfying the boundary
conditions.

\item If the star has a two-layer structure, we have to consider at $r
  \rightarrow 0$ the asymptotic expansion of the full set of equations
  (\ref{eqH1})--(\ref{eqcX}), in which the quantities $W^{lm}_{(b)}$,
  $H_1^{lm}$, $K^{lm}$, $X^{lm}$ are coupled with the superfluid degree of
  freedom $\delta\mu^{lm}$.  We find that the relations (\ref{ld1617}) remain
  unchanged [provided that one makes a replacement $W^{lm}(0) \rightarrow
  W_{(b)}^{lm}(0)$], while the expansion of the equation (\ref{ldmu}) at $r
  \rightarrow 0$, yields a new boundary condition:
\begin{eqnarray}
&&\delta\mu^{lm\,\prime\prime}(0)=
\frac{1}{2l+3}\left\{\frac{8\pi\rho_0}{3}l(l+2)\delta\mu^{lm}(0)-l\frac{h_2}{h_0}\delta
\mu^{lm}(0)\right.\nonumber\\
&&\left.+\frac{e^{-\nu_0/2}\omega^2}{h_0{\cal B}_0}\left[\delta\mu^{lm}(0)+\frac{\gamma_{2\,0}}{n_{b 0}
%\gamma_{3\,0}}{\tilde{\cal X}}^{lm}(0)\right]\right\}\,.\label{mu20}
\gamma_{3\,0}}{{\cal X}}^{lm}(0)\right]\right\}\,.\label{mu20}
\end{eqnarray} 
We note that since the differential equation for $\delta\mu^{lm}$ is of the
second order, in this case we need to include the second order term
$\delta\mu^{lm\,\prime\prime}(0)$ in the expansion.

Imposing the boundary conditions (\ref{ld1617}) and (\ref{mu20}) we have, for
each value of $\omega$, three independent solutions of the perturbation
equations (\ref{eqH1})-(\ref{eqcX}).
\end{itemize}
%

%%%%%%%%%%%%%%%%%%%%%%%%%%%%%%%%%%%%%%%%%%%%%%%%%%%%%%%%%%%%%%%%%%%%%%%%%%%%%%%%%%%%%%
\subsubsection{Outer boundary conditions}
%%%%%%%%%%%%%%%%%%%%%%%%%%%%%%%%%%%%%%%%%%%%%%%%%%%%%%%%%%%%%%%%%%%%%%%%%%%%%%%%%%%%%%
At the outer boundary ($r=r_f$) the oscillation equations imply
$W_{(b)}^{lm}(r_{f\,-})=W^{lm}(r_{f\,+})$, and continuity of $H_1^{lm}$,
$K^{lm}$, $X^{lm}$.  These conditions coincide with the corresponding boundary
conditions at $r=r_i$.  The situation with the boundary condition for the
quantity $\delta \mu^{lm}$ is more subtle.  If the superfluid phase does not
extend up to the crust, one has to impose Eq.~(\ref{bcrif}) at $r=r_f$.  If,
instead, the superfluid phase extends up to the crust, a boundary condition on
$\delta\mu^{lm}$ has to be imposed at the crust-core interface ($r_f=R_{\rm
  cc}$), where Eq.~(\ref{bcrif}) does not apply, because $h(r=R_{\rm cc}) \neq
0$.  In that case the appropriate boundary condition (see Ref.\ \cite{gkcg13})
follows from the requirement of the absence of particle transfer (baryons and
electrons) through the interface, that implies continuity of the radial velocity
$\delta u^r$ through the crust-core interface; this, combined with the condition
$W_{(b)}^{lm}(R_{cc\,-})=W^{lm}(R_{cc\,+})$, yields $X^r=0$ at $r=R_{cc}$. Using
Eq.\ (\ref{eq90}) the latter condition can be rewritten as \cite{cg11,gkcg13}
\begin{equation}
\left(\delta\mu^{lm\,\prime}+\frac{l}{r}\delta\mu^{lm}\right)_{r=R_{cc}}=0\,.\label{bcrc}
\end{equation}
The condition at the outer boundary of the superfluid phase [either
(\ref{bcrif}) or (\ref{bcrc})] reduces the number of independent solutions to
two.  We then integrate the standard LD equations, in terms of $W^{lm}$,
$V^{lm}$, etc., up to the NS surface, where we impose the vanishing of the
Lagrangian pressure perturbation, $X^{lm}(R)=0$.  After that only one solution
meeting all the boundary conditions inside the star survives.  Outside the star
we solve the Zerilli equation with two boundary conditions at the stellar
surface \cite{Zerilli:1971wd}. Finally, at infinity, we impose the vanishing of
the ingoing gravitational radiation.  This condition is satisfied by a discrete
set of (complex) frequencies $\omega$: the QNMs of the star.

%%%%%%%%%%%%%%%%%%%%%%%%%%%%%%%%%%%%%%%%%%%%%%%%%%%%%%%%%%%%%%%%%%%%%%%%%%%%%%%%%%%%%%
\section{Stellar models}\label{models}
%%%%%%%%%%%%%%%%%%%%%%%%%%%%%%%%%%%%%%%%%%%%%%%%%%%%%%%%%%%%%%%%%%%%%%%%%%%%%%%%%%%%%%
Microphysics input and equilibrium stellar models adopted in the present paper
are essentially the same as in Ref.\ \cite{gkcg13}.  We briefly describe them
here in order to make our presentation more self-contained.
%%%%%%%%%%%%%%%%%%%%%%%%%%%%%%%%%%%%%%%%%%%%%%%%%%%%%%%%%%%%%%%%%%%%%
\begin{figure}
    \begin{center}
        \leavevmode
        \epsfxsize=3.4in \epsfbox{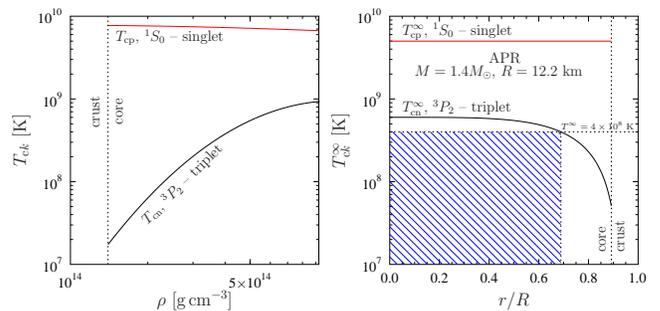}
    \end{center}
    \caption{(color online) Left panel: Nucleon critical temperatures
      $T_{{\mathrm c} k}$ ($k=n$, $p$) versus energy density $\rho$ for model A.
      Right panel: Redshifted critical temperatures $T^\infty_{{\mathrm c} k}$
      versus radial coordinate $r$ (in units of $R$) for model A.  }
    \label{Fig:Tc1}
\end{figure}
%%%%%%%%%%%%%%%%%%%%%%%%%%%%%%%%%%%%%%%%%%%%%%%%%%%%%%%%%%%%%%%%%%%%%

As mentioned in Sec.\ \ref{hydro}, we consider the simplest $npe$-matter
composition of NS core.  We adopt the Akmal-Pandharipande-Ravenhall EoS
\cite{apr98} parametrized in Ref.\ \cite{hh99} in the core and the equation of
state \cite{nv73} in the crust.

All numerical results presented here are obtained for a NS with mass
$M=1.4M_{\odot}$.  The circumferential radius for such star is $R=12.2$~km, the
central density is $\rho_{\rm c}=9.26 \times 10^{14}$~g~cm$^{-3}$.  We set the
crust-core interface at $\rho_{cc}=2\times 10^{14}$~g~cm$^{-3}$, at the distance
$R_{\rm cc}=10.9$~km from the centre.

We consider an isothermal temperature profile, i.e., we assume that the
redshifted temperature $T^\infty=e^{\nu/2}T$ is uniform over the core of the
star.  We also assume triplet pairing of neutrons and singlet pairing of
protons in the NS core.  The neutron superfluidity in the stellar crust is
ignored; this assumption should not noticeably affect global oscillations of
NSs.

Following Ref.\ \cite{gkcg13}, we consider two models of nucleon superfluidity,
which we denote by ``A'' and ``B'', as representatives of a two-layer and a
three-layer structure for the superfluid NS, respectively.

In model A the redshifted proton critical temperature is constant over the core,
$T_{{\rm c}p}^\infty \equiv T_{{\rm c}p} \, {\rm e}^{\nu/2}=5\times 10^9$~K; the
redshifted neutron critical temperature $T_{{\rm c}n}^\infty \equiv T_{{\rm c}n}
\, {\rm e}^{\nu/2}$ increases with the energy density $\rho$ and reaches the
maximum value $T_{{\rm c}n\, {\rm max}}^\infty=6\times 10^8$~K at the stellar
centre ($r$=0).  A similar model of neutron superfluidity (with the maximum of
$T_{cn}^\infty(\rho)$ at the stellar centre) has been recently considered in
Ref.\ \cite{bbbu13} and agrees with the results of some microscopic calculations
\cite{beehs98}.

In model B both critical temperatures $T_{{\rm c}n}^\infty$ and $T_{{\rm
    c}p}^\infty$ are density-dependent, and, depending on the value of the
temperature, the superfluid NS can have two or three layers. A similar model of
neutron superfluidity has been recently used to explain observations of the
cooling NS in Cassiopeia A supernova remnant \cite{Page:2010aw,syhhp11}, and
agrees with the results of microscopic calculations (see, e.g., Refs.\
\cite{ls01, yls99}).
%%%%%%%%%%%%%%%%%%%%%%%%%%%%%%%%%%%%%%%%%%%%%%%%%%%%%%%%%%%%%%%%%%%%%
\begin{figure}
   \begin{center}
       \leavevmode
        \epsfxsize=3.4in \epsfbox{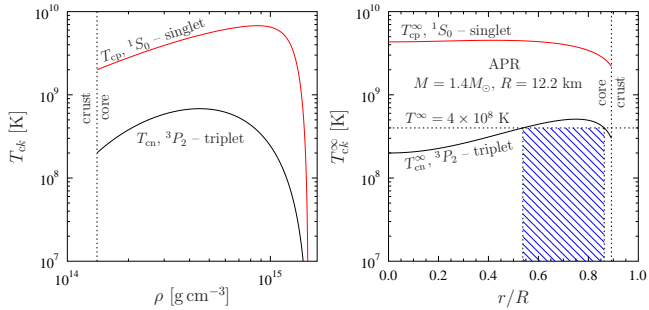}
    \end{center}
    \caption{(color online)
    Same as in Fig.\ \ref{Fig:Tc1}, for model B.
    }
    \label{Fig:Tc2}
\end{figure}
%%%%%%%%%%%%%%%%%%%%%%%%%%%%%%%%%%%%%%%%%%%%%%%%%%%%%%%%%%%%%%%%%%%%%

Models A and B are shown in Figs.~\ref{Fig:Tc1} and \ref{Fig:Tc2}.  These
figures coincide with, respectively, Figs.~1 and 2 of Ref. \cite{gkcg13}.  The
functions $T_{{\rm c}i}(\rho)$ are shown in the left panels of both figures; the
right panels show the dependence $T_{{\rm c}i}^\infty(r)$ ($i=n$ and $p$).  As
the redshifted temperature $T^\infty$ decreases, the size of the superfluid
region [given by the condition $T^\infty<T^\infty_{{\rm c}n}(r)$] increases or
remains unaffected.  For illustration, we shaded in Figs.\ \ref{Fig:Tc1} and
\ref{Fig:Tc2} the superfluid region corresponding to $T^\infty = 4\times10^8$~K.
One can see that in model B there can be three-layer configurations of a star
with no neutron superfluidity in the centre and in the outer region but with
superfluid intermediate region, or, for lower temperatures, two-layer
configurations.  In contrast, in model A only two-layer configurations are
possible.

%%%%%%%%%%%%%%%%%%%%%%%%%%%%%%%%%%%%%%%%%%%%%%%%%%%%%%%%%%%%%%%%%%%%%
\begin{figure}
   \begin{center}
\includegraphics[angle=0,width=6.cm]{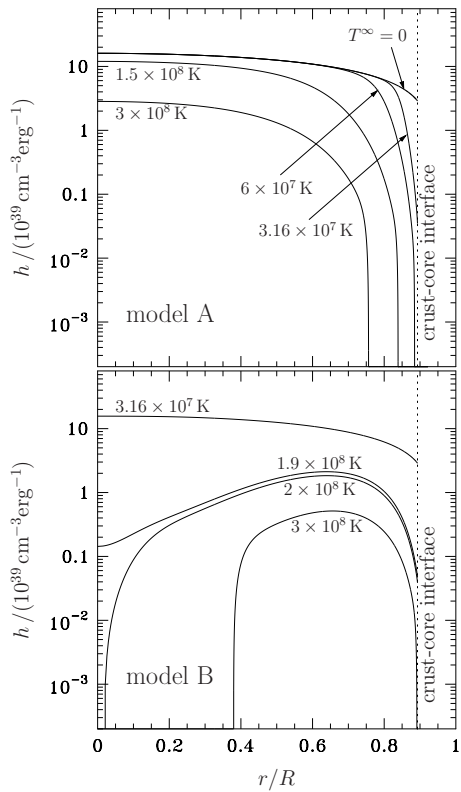}
\end{center}
\caption{Profiles of $h(r)$ for model A (upper panel) and model B (lower panel)
  for different values of stellar temperature (notice, that the zero temperature
  limit in model B practically coincides with the curve for $T^\infty=3.16\times
  10^7\,\rm K$). Vertical dotted line corresponds to the crust-core interface,
  $R_{cc}$.}\label{hplot}
\end{figure}
%%%%%%%%%%%%%%%%%%%%%%%%%%%%%%%%%%%%%%%%%%%%%%%%%%%%%%%%%%%%%%%%%%%%%
This can also be seen looking at the profiles of the function $h(r)$ defined in
Eq.~(\ref{defh}), which vanishes in the non-superfluid region, and is
non-vanishing in the superfluid region. In Fig.~\ref{hplot} we show $h(r)$ for
models $A$, $B$ and for different values of the temperature. We can see that
model $A$ yields two-layer configurations, while in model $B$ we have two-layer
configurations for $T^\infty\lesssim2\times 10^8$ K, and three-layer
configurations for $T^\infty\gtrsim2\times10^8$ K. For $T^\infty\ge6\times 10^8$
K (model A) or $T^\infty\gtrsim5\times 10^8$ K (model B), the superfluid region
disappears.

As we already emphasized in Sec.\ \ref{hydro}, the entrainment matrix $Y_{ik}$
depends on the critical temperature profiles $T_{ci}^\infty(\rho)$, and on the
value of the stellar temperature $T^\infty$ as well. We have computed $Y_{ik}$
for the models A and B following the same procedure as in Refs.\
\cite{gh05,gkh09b,gkcg13}.

%%%%%%%%%%%%%%%%%%%%%%%%%%%%%%%%%%%%%%%%%%%%%%%%%%%%%%%%%%%%%%%%%%%%%%%%%%%%%%%%%%%%%%
\section{Results}\label{results}
%%%%%%%%%%%%%%%%%%%%%%%%%%%%%%%%%%%%%%%%%%%%%%%%%%%%%%%%%%%%%%%%%%%%%%%%%%%%%%%%%%%%%%
Here we describe the results of our numerical integrations of the perturbative
equations derived in Sec.~\ref{gLD}, to find the QNMs of superfluid NSs. 

%%%%%%%%%%%%%%%%%%%%%%%%%%%%%%%%%%%%%%%%%%%%%%%%%%%%%%%%%%%%%%%%%%%%%%%%%%%%%%%%%%%%%%
\subsection{General structure of the QNM spectrum}{\label{genstruc}
%%%%%%%%%%%%%%%%%%%%%%%%%%%%%%%%%%%%%%%%%%%%%%%%%%%%%%%%%%%%%%%%%%%%%%%%%%%%%%%%%%%%%%
  As first noted by Epstein \cite{1988ApJ...333..880E} and Lindblom and Mendell
  \cite{1994ApJ...421..689L}, when a superfluid phase is present, there are two
  classes of QNMs.  The first class is formed by ``normal'' (or ``ordinary'')
  modes, which correspond (with small deviations) to modes of a non-superfluid
  NS. These modes, then, follow the standard classification (particularly, for
  $l\ge2$) in a fundamental mode (the $f$-mode), and a set of pressure modes
  (the $p_i$-modes) \cite{cox1980theory}.  The second class of modes is
  associated to the new degrees of freedom due to the relative motion of the
  fluids; they are called ``superfluid'' modes.  Notice, that superfluidity
  crucially affects the buoyancy modes, i.e., $g$-modes
  \cite{gk13th,Kantor:2014lja}, which cannot be classified neither as ``normal''
  modes, nor as ``superfluid'' modes.

  There is no standard notation for the superfluid modes. Some papers, such as
  \cite{1994ApJ...421..689L, 1995A&A...303..515L,Andersson:2002jd}, consider the
  superfluid modes as belonging to an unique class, and denote them as $s_i$, or
  $\beta_i$, etc. Other works \cite{Prix:2002fk,Wong:2008xa}, instead, follow
  the suggestion of \cite{Comer:1999rs}, where it was argued that a sort of
  doubling of the degrees of freedom occurs in a superfluid star, so there are
  two modes -- one ordinary and one superfluid -- for each $f$- or $p$- mode of
  a non-superfluid star. Therefore, there are the $f^o$- and the $f^s$-modes,
  the $p_i^o$- and the $p_i^s$-modes.  However, as noted in \cite{Prix:2002fk},
  this labeling is a pure convention, also because the number of nodes in the
  radial velocity eigenfunction does not always match with the order of the mode
  (in addition, it is not possible to define a single radial velocity
  eigenfunction in the superfluid case).
We then choose to treat the superfluid modes as part of an unique class, and
denote them as $s\!f_i$ ($i=0,1,\dots$) (we do not call them $s_i$ to avoid
confusion with purely gravitational modes with axial parity
\cite{1991RSPSA.434..449C}, which are also called $s_i$).

%%%%%%%%%%%%%%%%%%%%%%%%%%%%%%%%%%%%%%%%%%%%%%%%%%%%%%%%%%%%%%%%%%%%%%%%%%%%%%%%%%%%%%
\subsection{Comparison with the results of Refs.\ \cite{cg11,gkcg13} }
\label{compprev}
%%%%%%%%%%%%%%%%%%%%%%%%%%%%%%%%%%%%%%%%%%%%%%%%%%%%%%%%%%%%%%%%%%%%%%%%%%%%%%%%%%%%%%
In Refs.\ \cite{cg11,gkcg13} the spectrum of non-radial oscillations for
superfluid NSs was computed in the so called ``decoupled limit'', in which
equations governing the superfluid modes are completely decoupled from those
governing the normal (``ordinary'') modes.  As shown in Ref.\ \cite{gk11} this
approximation is very well justified, because the dimensionless parameter (the
coupling constant) $s=[n_e \, \partial P(n_b, \, n_e)/\partial n_e]/[n_b
\,\partial P(n_b, \, n_e)/\partial n_b]$, that couples superfluid and normal
degrees of freedom is actually small for realistic EoSs, $s \sim 0.01-0.05$.

In order to test our code, we have computed the frequencies of the superfluid
QNMs in the decoupled limit ($s=0$), and we have compared them with those
obtained in Refs.\ \cite{cg11,gkcg13}, using a completely different code.  We
have considered model $A$ and three values of the temperature: $T^\infty=0$
(cold star), $T^\infty=3.16\times10^7$ K (such that the superfluid phase fills
the whole core) and $T^\infty=6\times10^7$ K (at which the superfluid phase does
not fill the whole core).  We find [see Tab.\ \ref{comparison}, in which the
frequency is shown in units of $\tilde\nu=c/(2\pi R)$] a relative discrepancy
$\sim10^{-4}$, which we think can be explained in terms of the different
interpolation schemes which have been used.

%%%%%%%%%%%%%%%%%%%%%%%%%
\begin{table}
\begin{tabular}{|c|c|c|}
\hline
$T^\infty=0$ K&$T^\infty=3.16\times10^7$ K&$T^\infty=6\times10^7$ K\\
\hline
\begin{tabular}{cc}
(our code)& \cite{gkcg13}\\
\hline
$0.8309$&$0.8311$\\
$1.6137$&$1.6142$\\
$2.3166$&$2.3174$\\
\end{tabular}
&
\begin{tabular}{cc}
(our code)& \cite{gkcg13}\\
\hline
$0.8008$&$0.8011$\\
$1.5794$&$1.5799$\\
$2.2203$&$2.2211$\\
\end{tabular}
&
\begin{tabular}{cc}
(our code)& \cite{gkcg13}\\
\hline
$0.7088$&$0.7090$\\
$1.1220$&$1.1224$\\
$1.6364$&$1.6370$\\
\end{tabular}\\
\hline
\end{tabular}
\caption{Comparison between the frequencies of the first $l=2$ superfluid modes 
  ($s\!f_0$, $s\!f_1$, $s\!f_2$), computed in the decoupled case for model A, 
  using the  code developed for this work, and using the 
  code  employed in Ref.\ \cite{gkcg13}.  The frequencies are expressed in units
  of $c/(2\pi R)$, as in Ref.\ \cite{gkcg13}. \label{comparison}}
\end{table}
%%%%%%%%%%%%%%%%%%%%%%%%%%%%%%%%%%%%%%

%%%%%%%%%%%%%%%%%%%%%%%%%%%%%%%%%%%%%%%%%%%%%%%%%%%%%%%%%%%%%%%%%%%%%%%%%%%%%%%%%%%%%%
\subsection{The QNM spectrum of superfluid NSs}
\label{spect}
%%%%%%%%%%%%%%%%%%%%%%%%%%%%%%%%%%%%%%%%%%%%%%%%%%%%%%%%%%%%%%%%%%%%%%%%%%%%%%%%%%%%%%
We have computed the frequencies and gravitational damping times of the first
QNMs of the star, including the coupling between superfluid and non-superfluid
degrees of freedom, for models A and B. As expected (see Sec.~\ref{genstruc}) we
find two classes of QNMs: the normal $f$- and $p$-modes, and the superfluid
modes.

%%%%%%%%%%%%%%%%%%%%%%%%%%%%%%%%%%%%%%%%%%%%%%%%%%%%%%%%%%%%%%%%%%%%%%%%%%%%%%%%%%%%%%
\subsubsection{Frequencies}
%%%%%%%%%%%%%%%%%%%%%%%%%%%%%%%%%%%%%%%%%%%%%%%%%%%%%%%%%%%%%%%%%%%%%%%%%%%%%%%%%%%%%%
In Fig.~\ref{spectrumAB} we show the frequencies of the first $l=2$ QNMs as
functions of the redshifted temperature, for model A (upper panel) and B (lower
panel). In each panel we can see the first two normal modes (thin horizontal
lines), which are the same for models A and B and correspond to the frequencies
of a non-superfluid star with the same mass and EoS: the $f$-mode (at frequency
$\nu_f=1838$ Hz) and the $p_1$ mode (at frequency $\nu_{p_1}=5935$ Hz).  These
frequencies are generally not significantly affected by the presence of the
superfluid phase. Thus, the dependence of the normal-like modes on the
temperature is negligible.

Conversely, the superfluid modes strongly depend on the temperature, because it
determines the structure of the superfluid phase. In general, the frequency of a
superfluid mode decreases as the temperature increases, but in model B, at
temperatures close to $2\times10^8$ K, the behaviour is different. The reason is
that when the temperature becomes larger than $2\times10^8$ K, a phase
transition occurs, due to the appearance of a non-superfluid region at the
center of the star, and the structure of the superfluid NS changes from
two-layers to three-layers (see Sec.~\ref{models}). This transition is evident
in the lower panel of Fig.~\ref{spectrumAB}.  This behaviour was also evident in
the decoupled limit studied in \cite{cg11,gkcg13} (see, e.g., Fig.~6 of
\cite{gkcg13}).

Fig.~\ref{spectrumAB} also shows the occurrence of {\it avoided crossings}: at
particular values of the temperature (which we call {\it resonance temperatures}
$T_i^\infty$) the frequencies of some normal and superfluid modes become very
close, but the curves do not cross. A detail of the avoided crossing is shown in
the inset in upper panel of Fig.~\ref{spectrumAB} for model A.  This phenomenon
was expected, since it occurs in the case of radial pulsations
\cite{gk11,kg11,ga06}.  A similar phenomenon was also shown to occur, e.g., in
Refs.\ \cite{Prix:2002fk,Andersson:2002jd} (in non-rotating stars) and
\cite{Lee:2002fp,2003MNRAS.344..207Y} (for inertial modes of rotating stars),
studied in the zero-temperature limit. In these cases,
the frequencies of the modes were computed as functions of
the entrainment parameter, and it was shown that those curves had avoided
crossings.

%%%%%%%%%%%%%%%%%%%%%%%%%%%%%%%%%%%%%%%%%%%%%%%%%%%%%%%%%%%%%%%%%%%%%%%%%%%%%%%%%%%%%%
\begin{figure}[hbt]
\begin{center}
\includegraphics[angle=0,width=7.cm]{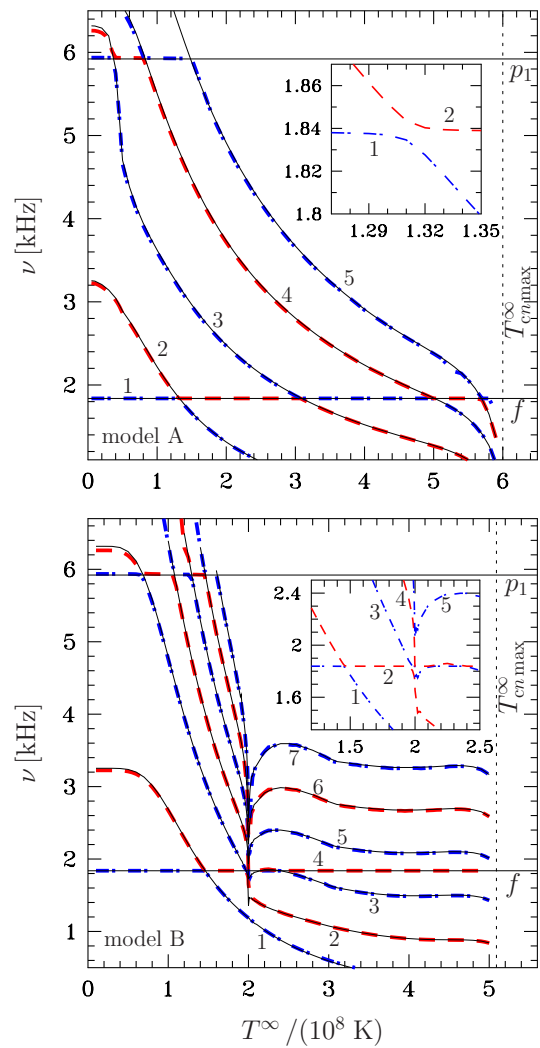}
\end{center}
\caption{(color online) Eigenfrequencies, $\nu$, of the first $l=2$ modes as
  functions of the redshifted stellar temperature $T^\infty$ for model A (upper
  panel) and model B (lower panel).  The oscillation modes (first 5 modes
  $1,\ldots,5$ for model A and first 7 modes $1,\ldots,7$ for model B) are shown
  by alternating dashed and dot-dashed lines.  The thin lines show superfluid
  and normal modes in the decoupled limit.  The inset in the upper panel shows
  one of the avoided crossings in detail.  The inset in the lower panel shows
  the first five modes near the phase transition temperature $T^\infty \approx 2
  \times 10^8\,\rm K$.  Vertical dotted lines indicate the maximum redshifted
  critical temperature for neutrons in the core ($T_{{\rm c}n\,\rm max}^\infty=6
  \times 10^8\,\rm K$ for model A, $T_{{\rm c}n\,\rm max}^\infty\approx 5.09
  \times 10^8\,\rm K$ for model B).}
\label{spectrumAB}
\end{figure}
%%%%%%%%%%%%%%%%%%%%%%%%%%%%%%%%%%%%%%%%%%%%%%%%%%%%%%%%%%%%%%%%%%%%%%%%%%%%%%%%%%%%%%

Finally, Fig.~\ref{spectrumAB} shows (thin solid lines) the frequencies of
superfluid and normal modes calculated in the decoupled limit.  It is clear that
the frequencies of the QNMs in the coupled and decoupled limits are very similar
for $T^\infty \neq T_i^\infty$. This is expected, since, as it was already noted
in Sec.\ \ref{compprev}, the coupling parameter $s$ is small for realistic EoSs
\cite{gk11}.  The coupling is crucial to determine the avoided crossings but,
far from the resonance temperatures $T_i^\infty$, the frequencies of the QNMs
are barely affected by the coupling. We can conclude that the approximation of
decoupled superfluid and normal modes works perfectly well for calculation of
the QNMs of superfluid NSs.

%%%%%%%%%%%%%%%%%%%%%%%%%%%%%%%%%%%%%%%%%%%%%%%%%%%%%%%%%%%%%%%%%%%%%%%%%%%%%%%%%%%%%%
\begin{figure}[hbt]
\begin{center}
\includegraphics[angle=0,width=6cm]{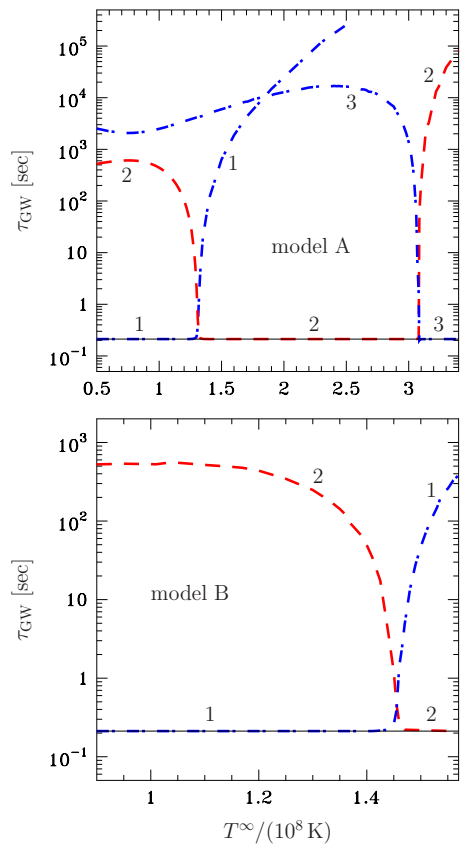}
\end{center}
\caption{(color online) Damping times for the lowest frequency modes shown in
  Fig.~\ref{spectrumAB} for model A (upper panel) and model B (lower panel) as
  functions of redshifted stellar temperature $T^\infty$.  Type of lines
  corresponds to that in Fig.~\ref{spectrumAB}.  Thin solid lines show the
  damping time of the $f$-mode in decoupled limit.}\label{damping}
\end{figure}
%%%%%%%%%%%%%%%%%%%%%%%%%%%%%%%%%%%%%%%%%%%%%%%%%%%%%%%%%%%%%%%%%%%%%%%%%%%%%%%%%%%%%%

%%%%%%%%%%%%%%%%%%%%%%%%%%%%%%%%%%%%%%%%%%%%%%%%%%%%%%%%%%%%%%%%%%%%%%%%%%%%%%%%%%%%%%
\subsubsection{Gravitational damping times}
%%%%%%%%%%%%%%%%%%%%%%%%%%%%%%%%%%%%%%%%%%%%%%%%%%%%%%%%%%%%%%%%%%%%%%%%%%%%%%%%%%%%%%
In Fig.~\ref{damping} we show the gravitational damping times $\tau_{\rm GW}$ of
the lowest frequency QNMs, as functions of redshifted temperature, for model A
(upper panel) and model B (lower panel).  In principle, our approach allows us
to compute $\tau_{\rm GW}$ for all of the QNMs\footnote{The approach described
  in Sec.~\ref{perturb} does not yield the viscous damping times, because
  dissipative terms are not included into the hydrodynamic equations. However,
  once the perturbation equations are solved, the viscous damping times can be
  computed, see, e.g., Ref.\ \cite{gkcg13}.}.  However, when the imaginary part
of the mode is much smaller than the real part, numerical errors make it
difficult to compute the damping times with good accuracy; this problem seems to
be more severe for temperatures $\lesssim5\times10^7$ K, and for damping times
$\gtrsim10^3-10^4$ s.  Still, we think that the values shown in
Fig.~\ref{damping} provide a reliable estimate at least of the order of
magnitude of the damping times, and of their dependence on the temperature.

We can see that at the resonance temperatures $T_i^\infty$, the curves of the
damping times do cross, and the modes change their nature from normal to
superfluid and vice versa.  Fig.~\ref{damping} also shows that, far from the
resonance temperatures $T_i^\infty$, the superfluid modes have damping times
$\gtrsim10^2-10^3$ s, much larger than those of the normal modes ($\sim 0.1-1$
s).  This result is consistent with calculations of Ref.\
\cite{Andersson:2002jd} and the prediction of Ref.\ \cite{cg11} (see also Ref.\
\cite{gk11}) that the intensity of the gravitational radiation should be
smaller, by a factor of $\sim s^2\simeq10^{-3}$, for superfluid modes than for
normal modes with similar frequencies.  Conversely, at temperatures close to
$T_i^\infty$, the damping times of the superfluid modes sharply decrease,
becoming comparable with those of the corresponding normal modes. This behaviour
is due to the fact that at $T^\infty\sim T_i^\infty$, the normal and superfluid
degrees of freedom become significantly coupled.  Thin lines in both panels of
Fig.~\ref{damping} show gravitational damping times $\tau_{\rm GW}$ for the
normal $f$-mode, which is calculated in the decoupled limit (notice that
$\tau_{\rm GW}=\infty$ for superfluid modes in this limit; thus they are not
shown here).

The viscous damping time for normal and superfluid modes $\tau_{\rm b+s}$, which
has been computed in Ref.\ \cite{gkcg13} in the decoupled limit, shows an
analogous qualitative behaviour. However, $\tau_{\rm b+s}$ for normal modes is
much larger than for superfluid modes, while for gravitational damping times the
situation is opposite.  Comparing our results with those of Ref.\ \cite{gkcg13},
we find that at moderate and high temperatures ($T^\infty \gtrsim 3 \times 10^7$
K) the viscous damping times for superfluid modes are significantly larger than
the gravitational damping times.  However, this comparison has been only made
for the lowest lying QNMs, because we have been able to compute $\tau_{\rm GW}$
for these modes only.  We note that, as the order of the mode increases, the
gravitational damping time increases, while the viscous damping time decreases
\cite{gkcg13}, therefore it is reasonable to expect that $\tau_{\rm GW}$ becomes
larger than $\tau_{\rm b+s}$ for high-order superfluid modes.  Moreover, even
low-order modes will be damped mostly due to (shear) viscosity if the stellar
temperature is sufficiently small.

QNMs with shorter damping times are more efficient in emitting GWs.  Indeed, the
GW flux can be estimated as $L_{\rm GW}\simeq 2E_{\rm mech}/\tau_{\rm GW}$
\cite{tc67}, where $E_{\rm mech}$ is the mechanical pulsation energy stored in
the mode, introduced in Sec.~\ref{Emech}.  Therefore, for generic values of the
temperature the superfluid modes are not good sources of GWs, because their
damping times are large; but, at temperatures close to the resonance
temperatures $T_i^\infty$, their damping times become comparable to those of the
normal modes, and they can become much more efficient in emitting GWs.  We can
expect, then, that at certain stages of NS thermal evolution, when a NS reaches
one of the resonance temperatures, a new QNM -- in principle detectable by GW
observers -- can appear in the GW spectrum.
%%%%%%%%%%%%%%%%%%%%%%%%%%%%%%%%%%%%%%%%%%%%%%%%%%%%%%%%%%%%%%%%%%%%%%%%%%%%%%%%%%%%%%
\subsubsection{Eigenfunctions}
%%%%%%%%%%%%%%%%%%%%%%%%%%%%%%%%%%%%%%%%%%%%%%%%%%%%%%%%%%%%%%%%%%%%%%%%%%%%%%%%%%%%%%

%%%%%%%%%%%%%%%%%%%%%%%%%%%%%%%%%%%%%%%%%%%%%%%%%%%%%%%%%%%%%%%%%%%%%%%%%%%%%%%%%%%%%%
\begin{figure}[hbt]
\begin{center}
\includegraphics[angle=0,width=6cm]{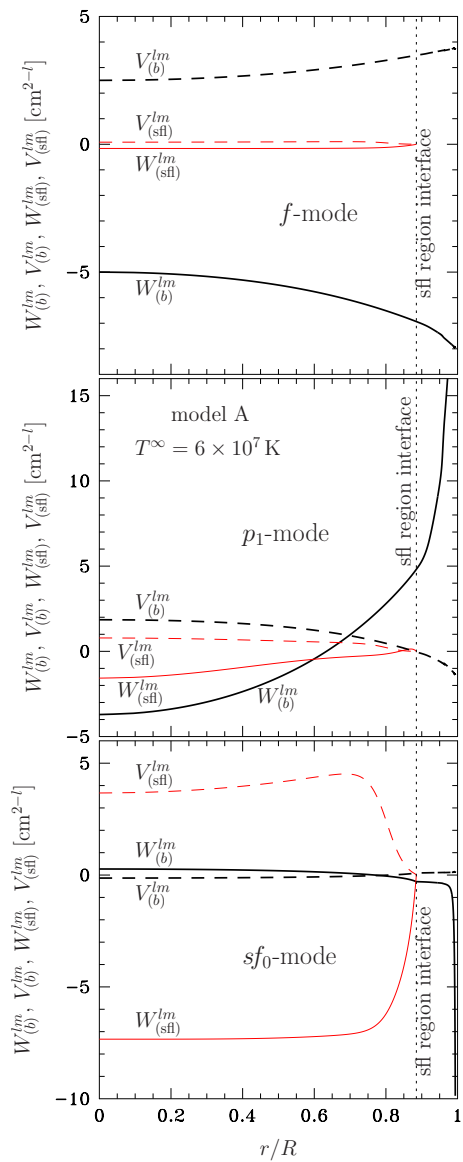}
\end{center}
\caption{(color online) Velocity eigenfunctions,
  $W^{lm}_{(b)},\,V^{lm}_{(b)},\,W^{lm}_{\rm (sfl)},\,V^{lm}_{\rm (sfl)}$, for
  the $l=2$ $f$-mode (upper panel), the $p_1$-mode (middle panel) and the
  $s\!f_0$-mode (lower panel) as functions of $r$ calculated in model A at
  $T^\infty=6\times10^7$ K. Vertical dotted lines correspond to the interface of
  the superfluid region.}\label{mods}
\end{figure}
%%%%%%%%%%%%%%%%%%%%%%%%%%%%%%%%%%%%%%%%%%%%%%%%%%%%%%%%%%%%%%%%%%%%%%%%%%%%%%%%%%%%%%

%%%%%%%%%%%%%%%%%%%%%%%%%%%%%%%%%%%%%%%%%%%%%%%%%%%%%%%%%%%%%%%%%%%%%%%%%%%%%%%%%%%%%%%
\begin{figure}[hbt]
\begin{center}
\includegraphics[angle=0,width=6cm]{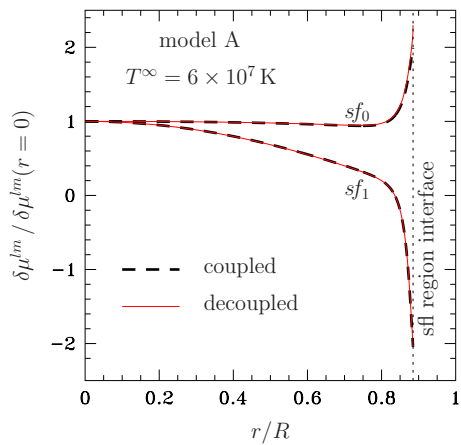}
\end{center}
\caption{(color online) Chemical potential imbalance eigenfunction, $\delta
  \mu^{lm}$, for the $l=2$ $s\!f_0$- and $s\!f_1$- modes as functions of $r$
  calculated in model A at $T^\infty=6\times10^7$ K.  Thin solid lines are
  calculated in the decoupled limit, while dashed lines are calculated including
  the coupling.  Vertical dotted line corresponds to the interface of the
  superfluid region.}\label{modmu}
\end{figure}
%%%%%%%%%%%%%%%%%%%%%%%%%%%%%%%%%%%%%%%%%%%%%%%%%%%%%%%%%%%%%%%%%%%%%%%%%%%%%%%%%%%%%%

In Fig.~\ref{mods} we show the velocity eigenfunctions for the $f$-mode (upper
panel), the $p_1$-mode (middle panel) and the $s\!f_0$-mode (lower panel), for
model A at $T^\infty=6\times10^7$ K.  We show the ($l=2$) quantities
$W^{lm}_{(b)}$, $V^{lm}_{(b)}$, obtained expanding the radial and angular
components, respectively, of the perturbation $\delta U_{(b)}^\mu$ in spherical
harmonics (\ref{expanUb}), and the quantities $W_{\rm(sfl)}^{lm}$,
$V_{\rm(sfl)}^{lm}$ obtained expanding in the same way $X^\mu$ [see Eq.\
(\ref{expanX})].  Note that the knowledge of these quantities allows one, using
Eq.\ (\ref{decdegr}), to calculate also the functions $W^{lm}$ and $V^{lm}$,
defined by the expansion of $\delta u^\mu$ (\ref{expanu}).

We can see that for the first superfluid mode $W^{lm}_{\rm(sfl)}\gg
W^{lm}_{(b)}$, $V^{lm}_{\rm(sfl)}\gg V^{lm}_{(b)}$.  This is a natural result,
since the coupling parameter $s$ is small and superfluid oscillations almost do
not excite baryon current (see Ref.\ \cite{gk11}). This supports the
interpretation (see Ref.\ \cite{gkcg13} and the footnote in Sec.~\ref{Emech}) of
$W^{lm}_{(b)}$, $V^{lm}_{(b)}$ as describing non-superfluid degrees of freedom
and $W^{lm}_{\rm(sfl)}$, $V^{lm}_{\rm(sfl)}$ as describing superfluid degrees of
freedom.  For the first pressure mode we obtain $W^{lm}_{(b)}\sim
W^{lm}_{\rm(sfl)}$, $V^{lm}_{(b)}\sim V^{lm}_{\rm(sfl)}$, while for the
fundamental mode $W^{lm}_{\rm(sfl)}\ll W^{lm}_{(b)}$, $V^{lm}_{\rm(sfl)}\ll
V^{lm}_{(b)}$.  The latter result is also expected and follows from the two
facts \cite{gkcg13}: ($i$) superfluid degrees of freedom (i.e., the quantities
$W^{lm}_{\rm(sfl)}$ and $V^{lm}_{\rm(sfl)}$) are excited by the gradient of the
chemical potential imbalance $\delta \mu$ [see Eqs.\ (\ref{expanX}) and
(\ref{eq90})] and ($ii$) $f$-mode oscillations are almost incompressible (i.e.,
deviation from the beta-equilibrium in the course of $f$-mode oscillations is
small), thus $\delta \mu$ is only weakly perturbed for $f$-modes.

From Fig.~\ref{mods} one can see that the radial
velocity eigenfunctions $W^{lm}_{(b)}$, $W^{lm}_{\rm (sfl)}$ have no nodes
inside the star in the case of the $f$-mode, one node in the case of the
$p_1$-mode. In the case of the $s\!f_0$ mode, the eigenfunction
$W^{lm}_{(b)}$ has one node, but $W^{lm}_{\rm (sfl)}$ (which is by far the
largest) has no nodes.

The eigenfunctions $\delta\mu^{lm}(r)$ for the $l=2$ $s\!f_0$ and $s\!f_1$ modes
calculated for model A and $T^\infty=6\times10^7$ K are shown in
Fig.~\ref{modmu} in the coupled (dashed lines) and decoupled (thin solid lines)
cases.  The very good agreement between the two solutions demonstrates the
accuracy of the decoupled limit.

%%%%%%%%%%%%%%%%%%%%%%%%%%%%%%%%%%%%%%%%%%%%%%%%%%%%%%%%%%%%%%%%%%%%%%%%%%%%%%%%%%%%%%
\subsubsection{Pulsation energy}
%%%%%%%%%%%%%%%%%%%%%%%%%%%%%%%%%%%%%%%%%%%%%%%%%%%%%%%%%%%%%%%%%%%%%%%%%%%%%%%%%%%%%%
We have computed the mechanical pulsation energy $E_{\rm mech}=E_{\rm
  mech\,(sfl)}+E_{{\rm mech}\,(b)}$ stored in the QNMs, using
Eqs.~(\ref{Emechb}) and (\ref{Emechsfl}).  In Table~\ref{energyratio} we show
the ratio $E_{\rm mech\,(sfl)}/E_{{\rm mech\,}(b)}$ for the $f$-mode, the
$p_1$-mode and the $s\!f_0$-mode, for models $A$ and $B$, at different values of
the redshifted temperature. We can see that when the star oscillates in a
non-superfluid mode, $E_{\rm mech\,(sfl)}\ll E_{{\rm mech\,}(b)}$, i.e., most of
the energy is stored in non-superfluid degrees of freedom (this is more evident
for the $f$-mode than for the $p_1$-mode).  When a NS oscillates in a superfluid
mode, $E_{\rm mech\,(sfl)}\gg E_{{\rm mech\,}(b)}$, i.e., most of the energy is
stored in superfluid degrees of freedom, while baryon currents are almost not
excited.

%%%%%%%%%%%%%%%%%%%%%%%%%%%%%%%%
\begin{table}
\begin{center}
\begin{tabular}{|c|c|ccc|}
\hline
Model&$T^\infty/(10^8\,\rm K)$&$f$&$p_1$&$s\!f_0$\\
\hline\hline
&$0.6$&$3.8\times10^{-5}$&$6.0\times10^{-3}$&$1.5\times10^2$\\
&$1.3$&$9.7\times10^{-2}$&$2.4\times10^{-3}$&$6.8~~~~~~~~$\\
$A$&$1.5$&$4.6\times10^{-4}$&$1.6\times10^{-3}$&$1.7\times10^2$\\
&$2.5$&$1.6\times10^{-5}$&$3.1\times10^{-4}$&$0.9\times10^2$\\
&$3.0$&$1.2\times10^{-4}$&$1.6\times10^{-4}$&$0.7\times10^2$\\
\hline
&$0.6$&$2.1\times10^{-5}$&$3.0\times10^{-2}$&$1.0\times10^2$\\
$B$&$1.5$&$4.5\times10^{-3}$&$1.6\times10^{-3}$&$0.9\times10^2$\\
&$3.0$&$6.1\times10^{-6}$&$1.5\times10^{-4}$&$0.2\times10^2$\\
\hline
\end{tabular}
\end{center}
\caption{\label{energyratio} Ratio $E_{\rm mech\,(sfl)}/E_{{\rm mech\,}(b)}$  
  for the $f$-mode, the $p_1$-mode and the $s\!f_0$-mode, for different 
  values of redshifted temperature, for the models $A$ and $B$. 
  Notice that this ratio is slightly amplified  for the $f$-mode 
  at $T^\infty=1.3 \times 10^8$~K in model A and
  at $T^\infty=1.5\times 10^8$ K in model B, 
  because these temperatures are close to the temperatures 
  of the avoided-crossings with superfluid modes, 
  see Fig.~\ref{spectrumAB}.}
\end{table}
%%%%%%%%%%%%%%%%%%%%%%%%%%%%%%%

%%%%%%%%%%%%%%%%%%%%%%%%%%%%%%%%%%%%%%%%%%%%%%%%%%%%%%%%%%%%%%%%%%%%%%%%%%%%%%%%%%%%%%
\section{Conclusions}
\label{concl}
%%%%%%%%%%%%%%%%%%%%%%%%%%%%%%%%%%%%%%%%%%%%%%%%%%%%%%%%%%%%%%%%%%%%%%%%%%%%%%%%%%%%%%

In this article we have derived the equations describing, in a general
relativistic framework, non-radial oscillations of non-rotating NSs with a
superfluid phase, including -- for the first time -- finite temperature effects.
We have numerically solved these equations, finding the QNMs of the NS.  We have
employed two different models of nucleon superfluidity, as representatives of a
two-layer and a three-layer structure, respectively;
similar models are currently used in the literature to explain astrophysical
observations \cite{Gusakov:2004se,Page:2004fy,bbbu13,Page:2010aw,syhhp11}.  

We find (as expected from previous results) two classes of modes: normal modes,
corresponding (with very minor differences) to the fundamental and pressure
modes of non-superfluid stars; and superfluid modes, directly associated to the
superfluid degrees of freedom.

The frequencies of normal modes are almost independent of the NS temperature,
but those of superfluid modes have a strong temperature dependence.  The curves
$\nu(T^\infty)$ of normal and superfluid modes show avoided crossings at
specific {\it resonance} values $T^\infty_i$ of the temperature.  Far from these values, the
frequencies of the modes are accurately described by the decoupling
approximation formulated in \cite{gk11} and studied in \cite{cg11,gkcg13}, where
the coupling between superfluid and non-superfluid degrees of freedom was
neglected; on the other hand, this coupling is important at temperatures close
to the resonance temperatures $T^\infty_i$.

Our approach allows to directly compute the gravitational damping times of the
QNMs.  We find (consistently with the results of Refs.\
\cite{gk11,cg11}) that the gravitational damping times of superfluid
modes are much larger ($\gtrsim10^2\!-\!10^3$ s) than those of normal modes, but
at the resonance temperatures they have a sharp decrease and become similar to
those of normal modes.\footnote{Analogous behaviour was noted in Ref.\
  \cite{gkcg13} for the viscous damping times $\tau_{\rm b+s}$ in the decoupled
  limit.  Viscous damping times for normal modes are much larger than for
  superfluid modes, and decrease sharply near the resonance temperatures.  We
  also note that for the lowest frequency modes $\tau_{\rm GW}<\tau_{\rm b+s}$
  at $T^\infty>3 \times 10^7\,\rm K$, therefore these modes are mainly damped by
  GW emission rather than by viscosity.}
These results imply that, when a NS, during its cooling, reaches one of the
resonance temperatures $T^\infty_i$, the superfluid modes become potentially
efficient GW sources, and may appear in the GW spectrum.

The fact that the frequencies of the QNMs as functions of the temperature show
avoided crossings confirms previous results for radial modes
\cite{ga06,kg11,gk11}, and suggests that $r$-modes of rotating, superfluid NSs
could have the same structure.  This would have far-reaching consequences, since
-- as shown in Refs.\ \cite{Gusakov:2013aza,Gusakov:2013jwa} -- such hypothesis
could allow to explain the puzzling observations of hot rapidly rotating NSs in
low-mass X-ray binaries.

%%%%%%%%%%%%%%%%%%%%%%%%%%%%%%%%%%%%%%%%%%%%%%%%%%%%%%%%%%%%%%%%%%%%%%%%%%%%%%
\begin{acknowledgments}
This work was partially supported 
by “NewCompStar” (COST Action MP1304), 
RFBR (grants 14-02-00868-a and 14-02-31616-mol-a), 
RF president programme 
(grants MK-506.2014.2 and NSh-294.2014.2) and
the Dynasty Foundation.
\end{acknowledgments}
%%%%%%%%%%%%%%%%%%%%%%%%%%%%%%%%%%%%%%%%%%%%%%%%%%%%%%%%%%%%%%%%%%%%%%%%%%%%%%

\appendix*

%%%%%%%%%%%%%%%%%%%%%%%%%%%%%%%%%%%%%%%%%%%%%%%%%%%%%%%%%%%%%%%%%%%%%%%%%%%%%%%%%%%%%%
\section{Explicit form of the perturbation equations}\label{equations}
%%%%%%%%%%%%%%%%%%%%%%%%%%%%%%%%%%%%%%%%%%%%%%%%%%%%%%%%%%%%%%%%%%%%%%%%%%%%%%%%%%%%%%

Non-radial perturbations of stationary, spherically symmetric, superfluid stars
are described in general relativity by four first-order differential equations
for the quantities $H_1^{lm}$, $K^{lm}$, $W^{lm}_{(b)}$, $X^{lm}$
(for brevity of notation we here omit the superscript $^{(0)}$ for background
quantities):
\begin{eqnarray}
 H_1^{lm\,\prime} &=&
 -\frac{1}{r} \biggl[ \ell+1+\frac{2me^\lambda}{r}+4\pi
  r^2e^\lambda(P-\rho) \biggr]H_1^{lm}
\nonumber\\&&
 + \frac{e^\lambda}{r}
 \left[ H_0^{lm} + K^{lm} - 16\pi(\rho+P)V^{lm}_{(b)} \right] \:,
\label{eqH1}\\
 K^{{lm}\,\prime} &=&
 \frac{1}{r} H_0^{lm} + \frac{\ell(\ell+1)}{2r} H_1^{lm}
\nonumber\\&&
 - \left[ \frac{\ell+1}{r}-\frac{\nu'}{2} \right] K^{lm} 
 - 8\pi(\rho+P)\frac{e^{\lambda/2}}{r} W^{lm}_{(b)} \:,
\label{eqK}\\
 W^{{lm}\,\prime}_{(b)} &=&
 - \frac{\ell+1}{r} W^{lm}_{(b)} 
 + re^{\lambda/2} \left[ \frac{e^{-\nu/2}}{\gamma P}{{\cal X}}^{lm} 
\right.\nonumber\\&&\left.
 - \frac{\ell(\ell+1)}{r^2} V^{lm}_{(b)} + \frac{1}{2}H_0^{lm} + K^{lm} \right] \:,
\label{eqW}\\
 X^{{lm}\,\prime} &=&
 -\frac{\ell}{r} X^{lm} + \frac{(\rho+P)e^{\nu/2}}{2}
 \Biggl\{ \left( \frac{1}{r}-\frac{\nu'}{2} \right)H_0^{lm}
\nonumber\\&&
 + \left[r\omega^2e^{-\nu} + \frac{\ell(\ell+1)}{2r}\right] H_1^{lm}
 + \left(\frac{3}{2}\nu' - \frac{1}{r}\right) K^{lm}
\nonumber\\&&
 - \frac{\ell(\ell+1)}{r^2}\nu' V^{lm}_{(b)} 
 - \frac{2}{r} 
 \Biggl[ 4\pi(\rho+P)e^{\lambda/2} + \omega^2e^{\lambda/2-\nu}
\nonumber\\&&
 - \frac{r^2}{2}
 \biggl(\frac{e^{-\lambda/2}}{r^2}\nu'\biggr)' \Biggr] W^{lm}_{(b)} \Biggr\}, 
\label{eqX}
\end{eqnarray}
and one second-order differential equation
\begin{eqnarray}
\delta\mu^{lm\,\prime\prime}=&-&\left[\frac{h'}{h}-\frac{\lambda'}{2}
+\frac{2(l+1)}{r}\right]\delta\mu^{lm\,\prime}\nonumber\\
&&-\left[(1-e^\lambda)\frac{l(l+1)}{r^2}+\frac{l}{r}\left(\frac{h'}{h}-
\frac{\lambda'}{2}\right)\right]\delta\mu^{lm}\nonumber\\
&&+e^{\lambda-\nu/2}\frac{\omega^2}{h{\cal B}}\left(\delta\mu^{lm}+
\frac{\gamma_2}{n_b\gamma_3}{{\cal X}}^{lm}\right)
\label{eqdmu}
\end{eqnarray}
for $\delta\mu^{lm}$. The quantities $H_0^{lm}$, $V^{lm}_{(b)}$, ${\cal X}^{lm}$ are
given by the algebraic relations:
\begin{eqnarray}
&&\left[ 3m + \frac{(\ell\!-\!1)(\ell\!+\!2)}{2}r + 4\pi r^3P \right] H_0^{lm}
- 8\pi r^3e^{-\nu/2} X^{lm}
\nonumber\\&&
+ \left[ \frac{\ell(\ell+1)}{2}(m+4\pi r^3P) 
  - \omega^2r^3e^{-(\lambda+\nu)} \right] H_1^{lm}
\nonumber\\&&
- \left[ \frac{(\ell-1)(\ell+2)}{2}r - \omega^2r^3e^{-\nu}
\right.\nonumber\\&&\left.
- \frac{e^\lambda}{r}(m+4\pi r^3P)(3m-r+4\pi r^3P) \right] K^{lm}=0 \:,
\label{eqH0}\\
&&\omega^2(\rho+P)e^{-\nu/2} V^{lm}_{(b)} = 
X^{lm} + \frac{P'}{r}e^{(\nu-\lambda)/2} W^{lm}_{(b)}
\nonumber\\&&
- \frac{e^{\nu/2}}{2}(\rho+P) H_0^{lm} \:,
\label{eqV}\\&&
{{\cal X}}^{lm}=\frac{1}{1-\gamma_2}
\left(X^{lm}+\gamma_3n_b\delta\mu^{lm}\right)\,.
\label{eqcX}
\end{eqnarray}
%
%%%%%%%%%%%%%%%%%%%%%%%%%%%%%%%%%%%%%%%%%%%%%%%%%%%%%%%%%%%%%%%%%%%%%%%%%%%%%%%%%%%%%%
\bibliography{sflref}

\begin{thebibliography}{70}
\expandafter\ifx\csname natexlab\endcsname\relax\def\natexlab#1{#1}\fi
\expandafter\ifx\csname bibnamefont\endcsname\relax
  \def\bibnamefont#1{#1}\fi
\expandafter\ifx\csname bibfnamefont\endcsname\relax
  \def\bibfnamefont#1{#1}\fi
\expandafter\ifx\csname citenamefont\endcsname\relax
  \def\citenamefont#1{#1}\fi
\expandafter\ifx\csname url\endcsname\relax
  \def\url#1{\texttt{#1}}\fi
\expandafter\ifx\csname urlprefix\endcsname\relax\def\urlprefix{URL }\fi
\providecommand{\bibinfo}[2]{#2}
\providecommand{\eprint}[2][]{\url{#2}}

\bibitem[{\citenamefont{Lattimer and Prakash}(2007)}]{Lattimer:2006xb}
\bibinfo{author}{\bibfnamefont{J.~M.} \bibnamefont{Lattimer}} \bibnamefont{and}
  \bibinfo{author}{\bibfnamefont{M.}~\bibnamefont{Prakash}},
  \bibinfo{journal}{Phys.Rept.} \textbf{\bibinfo{volume}{442}},
  \bibinfo{pages}{109} (\bibinfo{year}{2007}), \eprint{astro-ph/0612440}.

\bibitem[{\citenamefont{Andersson et~al.}(2011)\citenamefont{Andersson,
  Ferrari, Jones, Kokkotas, Krishnan et~al.}}]{Andersson:2009yt}
\bibinfo{author}{\bibfnamefont{N.}~\bibnamefont{Andersson}},
  \bibinfo{author}{\bibfnamefont{V.}~\bibnamefont{Ferrari}},
  \bibinfo{author}{\bibfnamefont{D.}~\bibnamefont{Jones}},
  \bibinfo{author}{\bibfnamefont{K.}~\bibnamefont{Kokkotas}},
  \bibinfo{author}{\bibfnamefont{B.}~\bibnamefont{Krishnan}},
  \bibnamefont{et~al.}, \bibinfo{journal}{Gen.Rel.Grav.}
  \textbf{\bibinfo{volume}{43}}, \bibinfo{pages}{409} (\bibinfo{year}{2011}),
  \eprint{0912.0384}.

\bibitem[{\citenamefont{{Andersson} et~al.}(2013)\citenamefont{{Andersson},
  {Baker}, {Belczynski}, {Bernuzzi}, {Berti}, {Cadonati},
  {Cerd{\'a}-Dur{\'a}n}, {Clark}, {Favata}, {Finn} et~al.}}]{anderssonetal13}
\bibinfo{author}{\bibfnamefont{N.}~\bibnamefont{{Andersson}}},
  \bibinfo{author}{\bibfnamefont{J.}~\bibnamefont{{Baker}}},
  \bibinfo{author}{\bibfnamefont{K.}~\bibnamefont{{Belczynski}}},
  \bibinfo{author}{\bibfnamefont{S.}~\bibnamefont{{Bernuzzi}}},
  \bibinfo{author}{\bibfnamefont{E.}~\bibnamefont{{Berti}}},
  \bibinfo{author}{\bibfnamefont{L.}~\bibnamefont{{Cadonati}}},
  \bibinfo{author}{\bibfnamefont{P.}~\bibnamefont{{Cerd{\'a}-Dur{\'a}n}}},
  \bibinfo{author}{\bibfnamefont{J.}~\bibnamefont{{Clark}}},
  \bibinfo{author}{\bibfnamefont{M.}~\bibnamefont{{Favata}}},
  \bibinfo{author}{\bibfnamefont{L.~S.} \bibnamefont{{Finn}}},
  \bibnamefont{et~al.}, \bibinfo{journal}{Classical and Quantum Gravity}
  \textbf{\bibinfo{volume}{30}}, \bibinfo{eid}{193002} (\bibinfo{year}{2013}),
  \eprint{1305.0816}.

\bibitem[{\citenamefont{Andersson and Kokkotas}(1998)}]{Andersson:1997rn}
\bibinfo{author}{\bibfnamefont{N.}~\bibnamefont{Andersson}} \bibnamefont{and}
  \bibinfo{author}{\bibfnamefont{K.~D.} \bibnamefont{Kokkotas}},
  \bibinfo{journal}{Mon.Not.Roy.Astron.Soc.} \textbf{\bibinfo{volume}{299}},
  \bibinfo{pages}{1059} (\bibinfo{year}{1998}), \eprint{gr-qc/9711088}.

\bibitem[{\citenamefont{Kokkotas et~al.}(2001)\citenamefont{Kokkotas,
  Apostolatos, and Andersson}}]{Kokkotas:1999mn}
\bibinfo{author}{\bibfnamefont{K.}~\bibnamefont{Kokkotas}},
  \bibinfo{author}{\bibfnamefont{T.}~\bibnamefont{Apostolatos}},
  \bibnamefont{and}
  \bibinfo{author}{\bibfnamefont{N.}~\bibnamefont{Andersson}},
  \bibinfo{journal}{Mon.Not.Roy.Astron.Soc.} \textbf{\bibinfo{volume}{320}},
  \bibinfo{pages}{307} (\bibinfo{year}{2001}), \eprint{gr-qc/9901072}.

\bibitem[{\citenamefont{Benhar et~al.}(2004)\citenamefont{Benhar, Ferrari, and
  Gualtieri}}]{Benhar:2004xg}
\bibinfo{author}{\bibfnamefont{O.}~\bibnamefont{Benhar}},
  \bibinfo{author}{\bibfnamefont{V.}~\bibnamefont{Ferrari}}, \bibnamefont{and}
  \bibinfo{author}{\bibfnamefont{L.}~\bibnamefont{Gualtieri}},
  \bibinfo{journal}{Phys.Rev.} \textbf{\bibinfo{volume}{D70}},
  \bibinfo{pages}{124015} (\bibinfo{year}{2004}), \eprint{astro-ph/0407529}.

\bibitem[{\citenamefont{Israel et~al.}(2005)\citenamefont{Israel, Belloni,
  Stella, Rephaeli, Gruber et~al.}}]{Israel:2005av}
\bibinfo{author}{\bibfnamefont{G.}~\bibnamefont{Israel}},
  \bibinfo{author}{\bibfnamefont{T.}~\bibnamefont{Belloni}},
  \bibinfo{author}{\bibfnamefont{L.}~\bibnamefont{Stella}},
  \bibinfo{author}{\bibfnamefont{Y.}~\bibnamefont{Rephaeli}},
  \bibinfo{author}{\bibfnamefont{D.}~\bibnamefont{Gruber}},
  \bibnamefont{et~al.}, \bibinfo{journal}{Astrophys.J.}
  \textbf{\bibinfo{volume}{628}}, \bibinfo{pages}{L53} (\bibinfo{year}{2005}),
  \eprint{astro-ph/0505255}.

\bibitem[{\citenamefont{Strohmayer and Watts}(2005)}]{Strohmayer:2005ks}
\bibinfo{author}{\bibfnamefont{T.~E.} \bibnamefont{Strohmayer}}
  \bibnamefont{and} \bibinfo{author}{\bibfnamefont{A.~L.} \bibnamefont{Watts}},
  \bibinfo{journal}{Astrophys.J.} \textbf{\bibinfo{volume}{632}},
  \bibinfo{pages}{L111} (\bibinfo{year}{2005}), \eprint{astro-ph/0508206}.

\bibitem[{\citenamefont{Strohmayer and Watts}(2006)}]{Strohmayer:2006py}
\bibinfo{author}{\bibfnamefont{T.~E.} \bibnamefont{Strohmayer}}
  \bibnamefont{and} \bibinfo{author}{\bibfnamefont{A.~L.} \bibnamefont{Watts}},
  \bibinfo{journal}{Astrophys.J.} \textbf{\bibinfo{volume}{653}},
  \bibinfo{pages}{593} (\bibinfo{year}{2006}), \eprint{astro-ph/0608463}.

\bibitem[{\citenamefont{Watts and Strohmayer}(2007)}]{Watts:2006mr}
\bibinfo{author}{\bibfnamefont{A.~L.} \bibnamefont{Watts}} \bibnamefont{and}
  \bibinfo{author}{\bibfnamefont{T.~E.} \bibnamefont{Strohmayer}},
  \bibinfo{journal}{Adv.Space Res.} \textbf{\bibinfo{volume}{40}},
  \bibinfo{pages}{1446} (\bibinfo{year}{2007}), \eprint{astro-ph/0612252}.

\bibitem[{\citenamefont{{Yakovlev} et~al.}(1999)\citenamefont{{Yakovlev},
  {Levenfish}, and {Shibanov}}}]{yls99}
\bibinfo{author}{\bibfnamefont{D.~G.} \bibnamefont{{Yakovlev}}},
  \bibinfo{author}{\bibfnamefont{K.~P.} \bibnamefont{{Levenfish}}},
  \bibnamefont{and} \bibinfo{author}{\bibfnamefont{Y.~A.}
  \bibnamefont{{Shibanov}}}, \bibinfo{journal}{Physics Uspekhi}
  \textbf{\bibinfo{volume}{42}}, \bibinfo{pages}{737} (\bibinfo{year}{1999}),
  \eprint{astro-ph/9906456}.

\bibitem[{\citenamefont{{Lombardo} and {Schulze}}(2001)}]{ls01}
\bibinfo{author}{\bibfnamefont{U.}~\bibnamefont{{Lombardo}}} \bibnamefont{and}
  \bibinfo{author}{\bibfnamefont{H.-J.} \bibnamefont{{Schulze}}}, in
  \emph{\bibinfo{booktitle}{Physics of Neutron Star Interiors}}, edited by
  \bibinfo{editor}{\bibfnamefont{D.}~\bibnamefont{{Blaschke}}},
  \bibinfo{editor}{\bibfnamefont{N.~K.} \bibnamefont{{Glendenning}}},
  \bibnamefont{and}
  \bibinfo{editor}{\bibfnamefont{A.}~\bibnamefont{{Sedrakian}}}
  (\bibinfo{year}{2001}), vol. \bibinfo{volume}{578} of
  \emph{\bibinfo{series}{Lecture Notes in Physics, Berlin Springer Verlag}},
  p.~\bibinfo{pages}{30}, \eprint{astro-ph/0012209}.

\bibitem[{\citenamefont{Chamel and Haensel}(2008)}]{lrr-2008-10}
\bibinfo{author}{\bibfnamefont{N.}~\bibnamefont{Chamel}} \bibnamefont{and}
  \bibinfo{author}{\bibfnamefont{P.}~\bibnamefont{Haensel}},
  \bibinfo{journal}{Living Reviews in Relativity} \textbf{\bibinfo{volume}{11}}
  (\bibinfo{year}{2008}),
  \urlprefix\url{http://www.livingreviews.org/lrr-2008-10}.

\bibitem[{\citenamefont{Heinke and Ho}(2010)}]{Heinke:2010cr}
\bibinfo{author}{\bibfnamefont{C.~O.} \bibnamefont{Heinke}} \bibnamefont{and}
  \bibinfo{author}{\bibfnamefont{W.~C.} \bibnamefont{Ho}},
  \bibinfo{journal}{Astrophys.J.} \textbf{\bibinfo{volume}{719}},
  \bibinfo{pages}{L167} (\bibinfo{year}{2010}), \eprint{1007.4719}.

\bibitem[{\citenamefont{Elshamouty et~al.}(2013)\citenamefont{Elshamouty,
  Heinke, Sivakoff, Ho, Shternin et~al.}}]{Elshamouty:2013nfa}
\bibinfo{author}{\bibfnamefont{K.}~\bibnamefont{Elshamouty}},
  \bibinfo{author}{\bibfnamefont{C.}~\bibnamefont{Heinke}},
  \bibinfo{author}{\bibfnamefont{G.}~\bibnamefont{Sivakoff}},
  \bibinfo{author}{\bibfnamefont{W.}~\bibnamefont{Ho}},
  \bibinfo{author}{\bibfnamefont{P.}~\bibnamefont{Shternin}},
  \bibnamefont{et~al.}, \bibinfo{journal}{Astrophys.J.}
  \textbf{\bibinfo{volume}{777}}, \bibinfo{pages}{22} (\bibinfo{year}{2013}),
  \eprint{1306.3387}.

\bibitem[{\citenamefont{{Shternin} et~al.}(2011)\citenamefont{{Shternin},
  {Yakovlev}, {Heinke}, {Ho}, and {Patnaude}}}]{syhhp11}
\bibinfo{author}{\bibfnamefont{P.~S.} \bibnamefont{{Shternin}}},
  \bibinfo{author}{\bibfnamefont{D.~G.} \bibnamefont{{Yakovlev}}},
  \bibinfo{author}{\bibfnamefont{C.~O.} \bibnamefont{{Heinke}}},
  \bibinfo{author}{\bibfnamefont{W.~C.~G.} \bibnamefont{{Ho}}},
  \bibnamefont{and} \bibinfo{author}{\bibfnamefont{D.~J.}
  \bibnamefont{{Patnaude}}}, \bibinfo{journal}{\mnras}
  \textbf{\bibinfo{volume}{412}}, \bibinfo{pages}{L108} (\bibinfo{year}{2011}),
  \eprint{1012.0045}.

\bibitem[{\citenamefont{Page et~al.}(2011)\citenamefont{Page, Prakash,
  Lattimer, and Steiner}}]{Page:2010aw}
\bibinfo{author}{\bibfnamefont{D.}~\bibnamefont{Page}},
  \bibinfo{author}{\bibfnamefont{M.}~\bibnamefont{Prakash}},
  \bibinfo{author}{\bibfnamefont{J.~M.} \bibnamefont{Lattimer}},
  \bibnamefont{and} \bibinfo{author}{\bibfnamefont{A.~W.}
  \bibnamefont{Steiner}}, \bibinfo{journal}{Phys.Rev.Lett.}
  \textbf{\bibinfo{volume}{106}}, \bibinfo{pages}{081101}
  (\bibinfo{year}{2011}), \eprint{1011.6142}.

\bibitem[{\citenamefont{{Thorne} and {Campolattaro}}(1967)}]{tc67}
\bibinfo{author}{\bibfnamefont{K.~S.} \bibnamefont{{Thorne}}} \bibnamefont{and}
  \bibinfo{author}{\bibfnamefont{A.}~\bibnamefont{{Campolattaro}}},
  \bibinfo{journal}{Astrophys. J.} \textbf{\bibinfo{volume}{149}},
  \bibinfo{pages}{591} (\bibinfo{year}{1967}).

\bibitem[{\citenamefont{Lindblom and Detweiler}(1983)}]{ld83}
\bibinfo{author}{\bibfnamefont{L.}~\bibnamefont{Lindblom}} \bibnamefont{and}
  \bibinfo{author}{\bibfnamefont{S.~L.} \bibnamefont{Detweiler}},
  \bibinfo{journal}{Astrophys. J. Suppl.} \textbf{\bibinfo{volume}{53}},
  \bibinfo{pages}{73} (\bibinfo{year}{1983}).

\bibitem[{\citenamefont{{Detweiler} and {Lindblom}}(1985)}]{dl85}
\bibinfo{author}{\bibfnamefont{S.}~\bibnamefont{{Detweiler}}} \bibnamefont{and}
  \bibinfo{author}{\bibfnamefont{L.}~\bibnamefont{{Lindblom}}},
  \bibinfo{journal}{Astrophys. J.} \textbf{\bibinfo{volume}{292}},
  \bibinfo{pages}{12} (\bibinfo{year}{1985}).

\bibitem[{\citenamefont{Chandrasekhar and Ferrari}(1991)}]{cf91}
\bibinfo{author}{\bibfnamefont{S.}~\bibnamefont{Chandrasekhar}}
  \bibnamefont{and} \bibinfo{author}{\bibfnamefont{V.}~\bibnamefont{Ferrari}},
  \bibinfo{journal}{Proc. Roy. Soc. Lond.} \textbf{\bibinfo{volume}{A432}},
  \bibinfo{pages}{247} (\bibinfo{year}{1991}).

\bibitem[{\citenamefont{{Epstein}}(1988)}]{1988ApJ...333..880E}
\bibinfo{author}{\bibfnamefont{R.~I.} \bibnamefont{{Epstein}}},
  \bibinfo{journal}{\apj} \textbf{\bibinfo{volume}{333}}, \bibinfo{pages}{880}
  (\bibinfo{year}{1988}).

\bibitem[{\citenamefont{{Lindblom} and {Mendell}}(1994)}]{1994ApJ...421..689L}
\bibinfo{author}{\bibfnamefont{L.}~\bibnamefont{{Lindblom}}} \bibnamefont{and}
  \bibinfo{author}{\bibfnamefont{G.}~\bibnamefont{{Mendell}}},
  \bibinfo{journal}{\apj} \textbf{\bibinfo{volume}{421}}, \bibinfo{pages}{689}
  (\bibinfo{year}{1994}).

\bibitem[{\citenamefont{{Lee}}(1995)}]{1995A&A...303..515L}
\bibinfo{author}{\bibfnamefont{U.}~\bibnamefont{{Lee}}},
  \bibinfo{journal}{\aap} \textbf{\bibinfo{volume}{303}}, \bibinfo{pages}{515}
  (\bibinfo{year}{1995}).

\bibitem[{\citenamefont{{Lindblom} and {Mendell}}(2000)}]{lm00}
\bibinfo{author}{\bibfnamefont{L.}~\bibnamefont{{Lindblom}}} \bibnamefont{and}
  \bibinfo{author}{\bibfnamefont{G.}~\bibnamefont{{Mendell}}},
  \bibinfo{journal}{\prd} \textbf{\bibinfo{volume}{61}}, \bibinfo{eid}{104003}
  (\bibinfo{year}{2000}), \eprint{gr-qc/9909084}.

\bibitem[{\citenamefont{Andersson and Comer}(2001)}]{Andersson:2001bz}
\bibinfo{author}{\bibfnamefont{N.}~\bibnamefont{Andersson}} \bibnamefont{and}
  \bibinfo{author}{\bibfnamefont{G.}~\bibnamefont{Comer}},
  \bibinfo{journal}{Mon.Not.Roy.Astron.Soc.} \textbf{\bibinfo{volume}{328}},
  \bibinfo{pages}{1129} (\bibinfo{year}{2001}), \eprint{astro-ph/0101193}.

\bibitem[{\citenamefont{Prix et~al.}(2002)\citenamefont{Prix, Comer, and
  Andersson}}]{Prix:2001xc}
\bibinfo{author}{\bibfnamefont{R.}~\bibnamefont{Prix}},
  \bibinfo{author}{\bibfnamefont{G.}~\bibnamefont{Comer}}, \bibnamefont{and}
  \bibinfo{author}{\bibfnamefont{N.}~\bibnamefont{Andersson}},
  \bibinfo{journal}{Astron.Astrophys.} \textbf{\bibinfo{volume}{381}},
  \bibinfo{pages}{178} (\bibinfo{year}{2002}), \eprint{astro-ph/0107176}.

\bibitem[{\citenamefont{Prix and Rieutord}(2002)}]{Prix:2002fk}
\bibinfo{author}{\bibfnamefont{R.}~\bibnamefont{Prix}} \bibnamefont{and}
  \bibinfo{author}{\bibfnamefont{M.~L.} \bibnamefont{Rieutord}},
  \bibinfo{journal}{Astron.Astrophys.} \textbf{\bibinfo{volume}{393}},
  \bibinfo{pages}{949} (\bibinfo{year}{2002}), \eprint{astro-ph/0204520}.

\bibitem[{\citenamefont{{Yoshida} and {Lee}}(2003)}]{YL03a}
\bibinfo{author}{\bibfnamefont{S.}~\bibnamefont{{Yoshida}}} \bibnamefont{and}
  \bibinfo{author}{\bibfnamefont{U.}~\bibnamefont{{Lee}}},
  \bibinfo{journal}{\mnras} \textbf{\bibinfo{volume}{344}},
  \bibinfo{pages}{207} (\bibinfo{year}{2003}), \eprint{astro-ph/0302313}.

\bibitem[{\citenamefont{Wong et~al.}(2009)\citenamefont{Wong, Lin, and
  Leung}}]{Wong:2008xa}
\bibinfo{author}{\bibfnamefont{K.}~\bibnamefont{Wong}},
  \bibinfo{author}{\bibfnamefont{L.}~\bibnamefont{Lin}}, \bibnamefont{and}
  \bibinfo{author}{\bibfnamefont{P.}~\bibnamefont{Leung}},
  \bibinfo{journal}{Astrophys.J.} \textbf{\bibinfo{volume}{699}},
  \bibinfo{pages}{1809} (\bibinfo{year}{2009}), \eprint{0812.3708}.

\bibitem[{\citenamefont{Andersson et~al.}(2009)\citenamefont{Andersson,
  Glampedakis, and Haskell}}]{Andersson:2008fg}
\bibinfo{author}{\bibfnamefont{N.}~\bibnamefont{Andersson}},
  \bibinfo{author}{\bibfnamefont{K.}~\bibnamefont{Glampedakis}},
  \bibnamefont{and} \bibinfo{author}{\bibfnamefont{B.}~\bibnamefont{Haskell}},
  \bibinfo{journal}{Phys.Rev.} \textbf{\bibinfo{volume}{D79}},
  \bibinfo{pages}{103009} (\bibinfo{year}{2009}), \eprint{0812.3023}.

\bibitem[{\citenamefont{{Haskell} et~al.}(2009)\citenamefont{{Haskell},
  {Andersson}, and {Passamonti}}}]{2009MNRAS.397.1464H}
\bibinfo{author}{\bibfnamefont{B.}~\bibnamefont{{Haskell}}},
  \bibinfo{author}{\bibfnamefont{N.}~\bibnamefont{{Andersson}}},
  \bibnamefont{and}
  \bibinfo{author}{\bibfnamefont{A.}~\bibnamefont{{Passamonti}}},
  \bibinfo{journal}{\mnras} \textbf{\bibinfo{volume}{397}},
  \bibinfo{pages}{1464} (\bibinfo{year}{2009}), \eprint{0902.1149}.

\bibitem[{\citenamefont{{Passamonti} et~al.}(2009)\citenamefont{{Passamonti},
  {Haskell}, and {Andersson}}}]{2009MNRAS.396..951P}
\bibinfo{author}{\bibfnamefont{A.}~\bibnamefont{{Passamonti}}},
  \bibinfo{author}{\bibfnamefont{B.}~\bibnamefont{{Haskell}}},
  \bibnamefont{and}
  \bibinfo{author}{\bibfnamefont{N.}~\bibnamefont{{Andersson}}},
  \bibinfo{journal}{\mnras} \textbf{\bibinfo{volume}{396}},
  \bibinfo{pages}{951} (\bibinfo{year}{2009}), \eprint{0812.3569}.

\bibitem[{\citenamefont{Samuelsson and Andersson}(2009)}]{Samuelsson:2009xz}
\bibinfo{author}{\bibfnamefont{L.}~\bibnamefont{Samuelsson}} \bibnamefont{and}
  \bibinfo{author}{\bibfnamefont{N.}~\bibnamefont{Andersson}},
  \bibinfo{journal}{Class.Quant.Grav.} \textbf{\bibinfo{volume}{26}},
  \bibinfo{pages}{155016} (\bibinfo{year}{2009}), \eprint{0903.2437}.

\bibitem[{\citenamefont{{Passamonti} and {Andersson}}(2011)}]{pa11}
\bibinfo{author}{\bibfnamefont{A.}~\bibnamefont{{Passamonti}}}
  \bibnamefont{and}
  \bibinfo{author}{\bibfnamefont{N.}~\bibnamefont{{Andersson}}},
  \bibinfo{journal}{\mnras} \textbf{\bibinfo{volume}{413}}, \bibinfo{pages}{47}
  (\bibinfo{year}{2011}), \eprint{1004.4563}.

\bibitem[{\citenamefont{{Passamonti} and {Andersson}}(2012)}]{pa12}
\bibinfo{author}{\bibfnamefont{A.}~\bibnamefont{{Passamonti}}}
  \bibnamefont{and}
  \bibinfo{author}{\bibfnamefont{N.}~\bibnamefont{{Andersson}}},
  \bibinfo{journal}{\mnras} \textbf{\bibinfo{volume}{419}},
  \bibinfo{pages}{638} (\bibinfo{year}{2012}), \eprint{1105.4787}.

\bibitem[{\citenamefont{Comer et~al.}(1999)\citenamefont{Comer, Langlois, and
  Lin}}]{Comer:1999rs}
\bibinfo{author}{\bibfnamefont{G.}~\bibnamefont{Comer}},
  \bibinfo{author}{\bibfnamefont{D.}~\bibnamefont{Langlois}}, \bibnamefont{and}
  \bibinfo{author}{\bibfnamefont{L.~M.} \bibnamefont{Lin}},
  \bibinfo{journal}{Phys.Rev.} \textbf{\bibinfo{volume}{D60}},
  \bibinfo{pages}{104025} (\bibinfo{year}{1999}), \eprint{gr-qc/9908040}.

\bibitem[{\citenamefont{Andersson et~al.}(2002)\citenamefont{Andersson, Comer,
  and Langlois}}]{Andersson:2002jd}
\bibinfo{author}{\bibfnamefont{N.}~\bibnamefont{Andersson}},
  \bibinfo{author}{\bibfnamefont{G.}~\bibnamefont{Comer}}, \bibnamefont{and}
  \bibinfo{author}{\bibfnamefont{D.}~\bibnamefont{Langlois}},
  \bibinfo{journal}{Phys.Rev.} \textbf{\bibinfo{volume}{D66}},
  \bibinfo{pages}{104002} (\bibinfo{year}{2002}), \eprint{gr-qc/0203039}.

\bibitem[{\citenamefont{Yoshida and Lee}(2003)}]{Yoshida:2003hc}
\bibinfo{author}{\bibfnamefont{S.}~\bibnamefont{Yoshida}} \bibnamefont{and}
  \bibinfo{author}{\bibfnamefont{U.}~\bibnamefont{Lee}},
  \bibinfo{journal}{Phys.Rev.} \textbf{\bibinfo{volume}{D67}},
  \bibinfo{pages}{124019} (\bibinfo{year}{2003}), \eprint{gr-qc/0304073}.

\bibitem[{\citenamefont{{Lin} et~al.}(2008)\citenamefont{{Lin}, {Andersson},
  and {Comer}}}]{lac08}
\bibinfo{author}{\bibfnamefont{L.-M.} \bibnamefont{{Lin}}},
  \bibinfo{author}{\bibfnamefont{N.}~\bibnamefont{{Andersson}}},
  \bibnamefont{and} \bibinfo{author}{\bibfnamefont{G.~L.}
  \bibnamefont{{Comer}}}, \bibinfo{journal}{\prd}
  \textbf{\bibinfo{volume}{78}}, \bibinfo{eid}{083008} (\bibinfo{year}{2008}),
  \eprint{0709.0660}.

\bibitem[{\citenamefont{Chugunov and Gusakov}(2011)}]{cg11}
\bibinfo{author}{\bibfnamefont{A.}~\bibnamefont{Chugunov}} \bibnamefont{and}
  \bibinfo{author}{\bibfnamefont{M.}~\bibnamefont{Gusakov}},
  \bibinfo{journal}{Mon. Not. Roy. Astron. Soc.}
  \textbf{\bibinfo{volume}{418}}, \bibinfo{pages}{L54} (\bibinfo{year}{2011}),
  \eprint{1107.4242}.

\bibitem[{\citenamefont{Gusakov
  et~al.}(2013{\natexlab{a}})\citenamefont{Gusakov, Kantor, Chugunov, and
  Gualtieri}}]{gkcg13}
\bibinfo{author}{\bibfnamefont{M.}~\bibnamefont{Gusakov}},
  \bibinfo{author}{\bibfnamefont{E.}~\bibnamefont{Kantor}},
  \bibinfo{author}{\bibfnamefont{A.}~\bibnamefont{Chugunov}}, \bibnamefont{and}
  \bibinfo{author}{\bibfnamefont{L.}~\bibnamefont{Gualtieri}},
  \bibinfo{journal}{Mon. Not. Roy. Astron. Soc.}
  \textbf{\bibinfo{volume}{428}}, \bibinfo{pages}{1518}
  (\bibinfo{year}{2013}{\natexlab{a}}), \eprint{1211.2452}.

\bibitem[{\citenamefont{Kantor and Gusakov}(2011)}]{kg11}
\bibinfo{author}{\bibfnamefont{E.~M.} \bibnamefont{Kantor}} \bibnamefont{and}
  \bibinfo{author}{\bibfnamefont{M.~E.} \bibnamefont{Gusakov}},
  \bibinfo{journal}{\prd} \textbf{\bibinfo{volume}{83}},
  \bibinfo{pages}{103008} (\bibinfo{year}{2011}).

\bibitem[{\citenamefont{Gusakov and Kantor}(2011)}]{gk11}
\bibinfo{author}{\bibfnamefont{M.~E.} \bibnamefont{Gusakov}} \bibnamefont{and}
  \bibinfo{author}{\bibfnamefont{E.~M.} \bibnamefont{Kantor}},
  \bibinfo{journal}{Phys. Rev.} \textbf{\bibinfo{volume}{D83}},
  \bibinfo{pages}{081304} (\bibinfo{year}{2011}), \eprint{1007.2752}.

\bibitem[{\citenamefont{Gusakov and Andersson}(2006)}]{ga06}
\bibinfo{author}{\bibfnamefont{M.}~\bibnamefont{Gusakov}} \bibnamefont{and}
  \bibinfo{author}{\bibfnamefont{N.}~\bibnamefont{Andersson}},
  \bibinfo{journal}{\mnras} \textbf{\bibinfo{volume}{372}},
  \bibinfo{pages}{1776} (\bibinfo{year}{2006}).

\bibitem[{\citenamefont{Gusakov}(2007)}]{gusakov07}
\bibinfo{author}{\bibfnamefont{M.~E.} \bibnamefont{Gusakov}},
  \bibinfo{journal}{\prd} \textbf{\bibinfo{volume}{76}},
  \bibinfo{pages}{083001} (\bibinfo{year}{2007}).

\bibitem[{\citenamefont{{Khalatnikov}}(2000)}]{khalatnikov89}
\bibinfo{author}{\bibfnamefont{I.}~\bibnamefont{{Khalatnikov}}},
  \emph{\bibinfo{title}{An Introduction to the Theory of Superfluidity}}
  (\bibinfo{year}{2000}).

\bibitem[{\citenamefont{Andreev and Bashkin}(1975)}]{ab75}
\bibinfo{author}{\bibfnamefont{A.}~\bibnamefont{Andreev}} \bibnamefont{and}
  \bibinfo{author}{\bibfnamefont{E.}~\bibnamefont{Bashkin}},
  \bibinfo{journal}{Zh. Eksp. Teor. Fiz., v. 69, no. 1, pp. 319-326}
  \textbf{\bibinfo{volume}{69}} (\bibinfo{year}{1975}).

\bibitem[{\citenamefont{Borumand et~al.}(1996)\citenamefont{Borumand, Joynt,
  and Klu{\'z}niak}}]{bjk96}
\bibinfo{author}{\bibfnamefont{M.}~\bibnamefont{Borumand}},
  \bibinfo{author}{\bibfnamefont{R.}~\bibnamefont{Joynt}}, \bibnamefont{and}
  \bibinfo{author}{\bibfnamefont{W.}~\bibnamefont{Klu{\'z}niak}},
  \bibinfo{journal}{\prc} \textbf{\bibinfo{volume}{54}}, \bibinfo{pages}{2745}
  (\bibinfo{year}{1996}).

\bibitem[{\citenamefont{{Gusakov} and {Haensel}}(2005)}]{gh05}
\bibinfo{author}{\bibfnamefont{M.~E.} \bibnamefont{{Gusakov}}}
  \bibnamefont{and}
  \bibinfo{author}{\bibfnamefont{P.}~\bibnamefont{{Haensel}}},
  \bibinfo{journal}{Nuclear Physics A} \textbf{\bibinfo{volume}{761}},
  \bibinfo{pages}{333} (\bibinfo{year}{2005}), \eprint{astro-ph/0508104}.

\bibitem[{\citenamefont{{Gusakov}}(2010)}]{gusakov10}
\bibinfo{author}{\bibfnamefont{M.~E.} \bibnamefont{{Gusakov}}},
  \bibinfo{journal}{\prc} \textbf{\bibinfo{volume}{81}}, \bibinfo{eid}{025804}
  (\bibinfo{year}{2010}), \eprint{1001.4452}.

\bibitem[{\citenamefont{Gusakov et~al.}(2009)\citenamefont{Gusakov, Kantor, and
  Haensel}}]{gkh09a}
\bibinfo{author}{\bibfnamefont{M.~E.} \bibnamefont{Gusakov}},
  \bibinfo{author}{\bibfnamefont{E.~M.} \bibnamefont{Kantor}},
  \bibnamefont{and} \bibinfo{author}{\bibfnamefont{P.}~\bibnamefont{Haensel}},
  \bibinfo{journal}{\prc} \textbf{\bibinfo{volume}{79}},
  \bibinfo{pages}{055806} (\bibinfo{year}{2009}).

\bibitem[{\citenamefont{{Gusakov} et~al.}(2009)\citenamefont{{Gusakov},
  {Kantor}, and {Haensel}}}]{gkh09b}
\bibinfo{author}{\bibfnamefont{M.~E.} \bibnamefont{{Gusakov}}},
  \bibinfo{author}{\bibfnamefont{E.~M.} \bibnamefont{{Kantor}}},
  \bibnamefont{and}
  \bibinfo{author}{\bibfnamefont{P.}~\bibnamefont{{Haensel}}},
  \bibinfo{journal}{\prc} \textbf{\bibinfo{volume}{80}}, \bibinfo{eid}{015803}
  (\bibinfo{year}{2009}), \eprint{0907.0010}.

\bibitem[{\citenamefont{{Gusakov} et~al.}(2014)\citenamefont{{Gusakov},
  {Haensel}, and {Kantor}}}]{ghk14}
\bibinfo{author}{\bibfnamefont{M.~E.} \bibnamefont{{Gusakov}}},
  \bibinfo{author}{\bibfnamefont{P.}~\bibnamefont{{Haensel}}},
  \bibnamefont{and} \bibinfo{author}{\bibfnamefont{E.~M.}
  \bibnamefont{{Kantor}}}, \bibinfo{journal}{\mnras}
  \textbf{\bibinfo{volume}{439}}, \bibinfo{pages}{318} (\bibinfo{year}{2014}),
  \eprint{1401.2827}.

\bibitem[{\citenamefont{Zerilli}(1970)}]{Zerilli:1971wd}
\bibinfo{author}{\bibfnamefont{F.}~\bibnamefont{Zerilli}},
  \bibinfo{journal}{Phys.Rev.} \textbf{\bibinfo{volume}{D2}},
  \bibinfo{pages}{2141} (\bibinfo{year}{1970}).

\bibitem[{\citenamefont{{Akmal} et~al.}(1998)\citenamefont{{Akmal},
  {Pandharipande}, and {Ravenhall}}}]{apr98}
\bibinfo{author}{\bibfnamefont{A.}~\bibnamefont{{Akmal}}},
  \bibinfo{author}{\bibfnamefont{V.~R.} \bibnamefont{{Pandharipande}}},
  \bibnamefont{and} \bibinfo{author}{\bibfnamefont{D.~G.}
  \bibnamefont{{Ravenhall}}}, \bibinfo{journal}{\prc}
  \textbf{\bibinfo{volume}{58}}, \bibinfo{pages}{1804} (\bibinfo{year}{1998}),
  \eprint{hep-ph/9804388}.

\bibitem[{\citenamefont{{Heiselberg} and {Hjorth-Jensen}}(1999)}]{hh99}
\bibinfo{author}{\bibfnamefont{H.}~\bibnamefont{{Heiselberg}}}
  \bibnamefont{and}
  \bibinfo{author}{\bibfnamefont{M.}~\bibnamefont{{Hjorth-Jensen}}},
  \bibinfo{journal}{\apjl} \textbf{\bibinfo{volume}{525}}, \bibinfo{pages}{L45}
  (\bibinfo{year}{1999}), \eprint{astro-ph/9904214}.

\bibitem[{\citenamefont{{Negele} and {Vautherin}}(1973)}]{nv73}
\bibinfo{author}{\bibfnamefont{J.~W.} \bibnamefont{{Negele}}} \bibnamefont{and}
  \bibinfo{author}{\bibfnamefont{D.}~\bibnamefont{{Vautherin}}},
  \bibinfo{journal}{Nucl.\ Phys.\ A} \textbf{\bibinfo{volume}{207}},
  \bibinfo{pages}{298} (\bibinfo{year}{1973}).

\bibitem[{\citenamefont{{Bonanno} et~al.}(2013)\citenamefont{{Bonanno},
  {Baldo}, {Burgio}, and {Urpin}}}]{bbbu13}
\bibinfo{author}{\bibfnamefont{A.}~\bibnamefont{{Bonanno}}},
  \bibinfo{author}{\bibfnamefont{M.}~\bibnamefont{{Baldo}}},
  \bibinfo{author}{\bibfnamefont{G.~F.} \bibnamefont{{Burgio}}},
  \bibnamefont{and} \bibinfo{author}{\bibfnamefont{V.}~\bibnamefont{{Urpin}}},
  \bibinfo{journal}{ArXiv e-prints}  (\bibinfo{year}{2013}),
  \eprint{1311.2153}.

\bibitem[{\citenamefont{{Baldo} et~al.}(1998)\citenamefont{{Baldo},
  {Elgar{\o}y}, {Engvik}, {Hjorth-Jensen}, and {Schulze}}}]{beehs98}
\bibinfo{author}{\bibfnamefont{M.}~\bibnamefont{{Baldo}}},
  \bibinfo{author}{\bibfnamefont{{\O}.}~\bibnamefont{{Elgar{\o}y}}},
  \bibinfo{author}{\bibfnamefont{L.}~\bibnamefont{{Engvik}}},
  \bibinfo{author}{\bibfnamefont{M.}~\bibnamefont{{Hjorth-Jensen}}},
  \bibnamefont{and} \bibinfo{author}{\bibfnamefont{H.-J.}
  \bibnamefont{{Schulze}}}, \bibinfo{journal}{\prc}
  \textbf{\bibinfo{volume}{58}}, \bibinfo{pages}{1921} (\bibinfo{year}{1998}),
  \eprint{nucl-th/9806097}.

\bibitem[{\citenamefont{Cox}(1980)}]{cox1980theory}
\bibinfo{author}{\bibfnamefont{J.~P.} \bibnamefont{Cox}},
  \bibinfo{journal}{Research supported by the National Science Foundation
  Princeton, NJ, Princeton University Press, 1980. 393 p.}
  \textbf{\bibinfo{volume}{1}} (\bibinfo{year}{1980}).

\bibitem[{\citenamefont{{Gusakov} and {Kantor}}(2013)}]{gk13th}
\bibinfo{author}{\bibfnamefont{M.~E.} \bibnamefont{{Gusakov}}}
  \bibnamefont{and} \bibinfo{author}{\bibfnamefont{E.~M.}
  \bibnamefont{{Kantor}}}, \bibinfo{journal}{\prd}
  \textbf{\bibinfo{volume}{88}}, \bibinfo{eid}{101302} (\bibinfo{year}{2013}).

\bibitem[{\citenamefont{Kantor and Gusakov}(2014)}]{Kantor:2014lja}
\bibinfo{author}{\bibfnamefont{E.}~\bibnamefont{Kantor}} \bibnamefont{and}
  \bibinfo{author}{\bibfnamefont{M.}~\bibnamefont{Gusakov}}
  (\bibinfo{year}{2014}), \eprint{1404.6768}.

\bibitem[{\citenamefont{{Chandrasekhar} and
  {Ferrari}}(1991)}]{1991RSPSA.434..449C}
\bibinfo{author}{\bibfnamefont{S.}~\bibnamefont{{Chandrasekhar}}}
  \bibnamefont{and}
  \bibinfo{author}{\bibfnamefont{V.}~\bibnamefont{{Ferrari}}},
  \bibinfo{journal}{Royal Society of London Proceedings Series A}
  \textbf{\bibinfo{volume}{434}}, \bibinfo{pages}{449} (\bibinfo{year}{1991}).

\bibitem[{\citenamefont{Lee and Yoshida}(2003)}]{Lee:2002fp}
\bibinfo{author}{\bibfnamefont{U.}~\bibnamefont{Lee}} \bibnamefont{and}
  \bibinfo{author}{\bibfnamefont{S.}~\bibnamefont{Yoshida}},
  \bibinfo{journal}{Astrophys.J.} \textbf{\bibinfo{volume}{586}},
  \bibinfo{pages}{403} (\bibinfo{year}{2003}), \eprint{astro-ph/0211580}.

\bibitem[{\citenamefont{{Yoshida} and {Lee}}(2003)}]{2003MNRAS.344..207Y}
\bibinfo{author}{\bibfnamefont{S.}~\bibnamefont{{Yoshida}}} \bibnamefont{and}
  \bibinfo{author}{\bibfnamefont{U.}~\bibnamefont{{Lee}}},
  \bibinfo{journal}{Mon.Not.Roy.Astron.Soc.} \textbf{\bibinfo{volume}{344}},
  \bibinfo{pages}{207} (\bibinfo{year}{2003}), \eprint{astro-ph/0302313}.

\bibitem[{\citenamefont{Gusakov et~al.}(2004)\citenamefont{Gusakov, Kaminker,
  Yakovlev, and Gnedin}}]{Gusakov:2004se}
\bibinfo{author}{\bibfnamefont{M.}~\bibnamefont{Gusakov}},
  \bibinfo{author}{\bibfnamefont{A.~D.} \bibnamefont{Kaminker}},
  \bibinfo{author}{\bibfnamefont{D.}~\bibnamefont{Yakovlev}}, \bibnamefont{and}
  \bibinfo{author}{\bibfnamefont{O.~Y.} \bibnamefont{Gnedin}},
  \bibinfo{journal}{Astron.Astrophys.} \textbf{\bibinfo{volume}{423}},
  \bibinfo{pages}{1063} (\bibinfo{year}{2004}), \eprint{astro-ph/0404002}.

\bibitem[{\citenamefont{Page et~al.}(2004)\citenamefont{Page, Lattimer,
  Prakash, and Steiner}}]{Page:2004fy}
\bibinfo{author}{\bibfnamefont{D.}~\bibnamefont{Page}},
  \bibinfo{author}{\bibfnamefont{J.~M.} \bibnamefont{Lattimer}},
  \bibinfo{author}{\bibfnamefont{M.}~\bibnamefont{Prakash}}, \bibnamefont{and}
  \bibinfo{author}{\bibfnamefont{A.~W.} \bibnamefont{Steiner}},
  \bibinfo{journal}{Astrophys.J.Suppl.} \textbf{\bibinfo{volume}{155}},
  \bibinfo{pages}{623} (\bibinfo{year}{2004}), \eprint{astro-ph/0403657}.

\bibitem[{\citenamefont{Gusakov
  et~al.}(2013{\natexlab{b}})\citenamefont{Gusakov, Chugunov, and
  Kantor}}]{Gusakov:2013aza}
\bibinfo{author}{\bibfnamefont{M.~E.} \bibnamefont{Gusakov}},
  \bibinfo{author}{\bibfnamefont{A.~I.} \bibnamefont{Chugunov}},
  \bibnamefont{and} \bibinfo{author}{\bibfnamefont{E.~M.} \bibnamefont{Kantor}}
  (\bibinfo{year}{2013}{\natexlab{b}}), \eprint{1305.3825}.

\bibitem[{\citenamefont{Gusakov et~al.}(2014)\citenamefont{Gusakov, Chugunov,
  and Kantor}}]{Gusakov:2013jwa}
\bibinfo{author}{\bibfnamefont{M.~E.} \bibnamefont{Gusakov}},
  \bibinfo{author}{\bibfnamefont{A.~I.} \bibnamefont{Chugunov}},
  \bibnamefont{and} \bibinfo{author}{\bibfnamefont{E.~M.}
  \bibnamefont{Kantor}}, \bibinfo{journal}{Phys.Rev.Lett.}
  \textbf{\bibinfo{volume}{112}}, \bibinfo{pages}{151101}
  (\bibinfo{year}{2014}), \eprint{1310.8103}.

\end{thebibliography}

\end{document}